# Identification of non-local continua for lattice-like materials

Andrea Bacigalupo and Luigi Gambarotta*

Department of Civil, Chemical and Environmental Engineering, University of Genova, Italy

**Abstract**

The paper is focused on the dynamic homogenization of lattice-like materials with lumped mass at the nodes to obtain energetically consistent models providing accurate descriptions of the acoustic behavior of the discrete system. The equation of motion of the Lagrangian one-dimensional lattice is transformed according to a unitary approach aimed to identify equivalent non-local continuum models of integral-differential and gradient type, the latter obtained through standard or enhanced continualization. The bilateral Z-transform of the difference equation of motion of the lattice, mapped on the unit circle, is matched to the governing integral-differential equation of the equivalent continuum in the transformed Fourier space, which has the same frequency band structure as the Lagrangian one. Firstly, it is shown that the approximation of the kernels via Taylor polynomials leads to the differential field equations of higher order continua endowed with non-local constitutive terms. The field equations derived from such approach corresponds to the ones obtained through the so called standard continualization. However, the differential problem turns out to be ill-posed because the non-positive definiteness of the potential energy density of the higher order continuum. Energetically consistent equivalent continua have been identified through a proper mapping correlating the transformed macro-displacements in the Fourier space with a new auxiliary regularizing continuum macro-displacement field in the same space. Specifically, the mapping here introduced has zeros at the edge of the first Brillouin zone. The integral-differential governing equation and the corresponding differential one has been reformulated through an enhanced continualization, that is characterized by energetically consistent differential equations with inertial and constitutive non-localities. Here, the constitutive and inertial kernels of the integral-differential equation exhibit polar singularities at the edge of the first Brillouin zone. The proposed approach is generalized in a consistent way to two-dimensional lattices by using multidimensional Z- and Fourier transforms, a procedure that may be easily extended to three-dimensional lattices. Finally, two examples of lattice-like systems consisting of periodic pre-stressed cable-nets of point mass at the nodes are analyzed. The resulting gradient continuum models provide dispersion curves who are in excellent agreement with those of the Lagrangian systems.

**Keywords:** Lattice-like materials; Dispersive wave propagation; Homogenization; Non-local continua; Bloch spectra.

---

* Corresponding Author

## 1. Introduction

Lattice materials like cellular, porous, reticulated systems with regular or stochastic structure are a class of functional-structural materials whose physical and mechanical overall properties may be tailoring designed by controlling the morphology of the micro-structure and its constitutive properties (see for instance Deshpande *et al.*, 2001, Martinsson and Movchan, 2003, Fleck *et al.*, 2010). This circumstance allows for large-scale production of truss and beam periodic reticulated lattices endowed with remarkable static and dynamic properties also through innovative three-dimensional printing and additive manufacturing technology. Recently, innovative applications of periodic lattice materials are focused to the active and passive control of the mechanical behavior through different techniques including time modulated, auxetic tunable lattices and others devices for the frequency spectral tuning (see for instance Wang *et al.*, 2018, Chen *et al.*, 2017, Chen *et al.*, 2019, Kadic *et al.*, 2019, Carta *et al.,* 2019 among the others).

The mechanics of such inhomogeneous materials, i.e. *i)* stiffness and strength (see Hutchinson and Fleck, 2006), *ii)* size effects (see Mindlin, 1964), *iii)* dispersive elastic wave propagation (see Brillouin, 1946) and others, may be caught through discrete lattice models which result very effective and in some cases these models may provide analytic results because their simplicity (see Colquitt *et al.*, 2011). However, when large complex systems with a great number of degrees of freedom have to be analysed or when a synthetic description of the mechanical response is required, as in the study of acoustic wave propagation, discrete lattice models are precluded and equivalent continuum models have to be identified (see Kunin, 1982, Triantafyllidis and Bardenhagen, 1993, Kevrekidis *et al.*, 2002).

Among the different homogenization techniques of periodic lattices, also known as *continualization methods*, the first-order homogenization is the simplest one. A local equivalent continuum is identified by replacing the discrete equations of motion with differential governing equations. In this case, the nodal displacements in the governing equation of the discrete model are replaced by the first order Taylor approximation of a continuum displacement field defined on an equivalent domain. Analogous results may be obtained by imposing the energy macro-homogeneity condition between the discrete and the continuum models. On the other hand, it is known that these classical models are not able to catch the size effects associated to the microstructure length scale (Mindlin, 1964) as well as to describe the acoustic wave dispersion occurring in the discrete periodic material as a consequence of Bragg and Mie scattering (Lu *et al.*, 2009). These drawbacks may be circumvented by exploiting *generalized continualization techniques* to identify non-local equivalent continua.

The approach based on the *continualization of the Lagrangian functional* of the lattice relies on an approximation of the difference of displacements of adjacent nodes of the discrete model through a truncated Taylor expansion of the continuum macro-displacement field. The elastic potential and kinetic energy of the equivalent continuum are obtained and hence the Euler-Lagrange equation is derived from the Lagrangian functional (see Metrikine and Askes, 2002, Kumar and McDowell, 2004, Askes and Metrikine, 2005, Polyzos and Fotiadis, 2012, Bacigalupo and Gambarotta, 2017a, among the others). If a second order expansion of the macro-displacement field is considered, an equivalent continuum governed by fourth order differential equations of motion is obtained. Similar results have been achieved by exploiting the shift operators and by introducing the concept of pseudo-differential operators and their approximations through power series (Rosenau, 2003). While the potential energy density of the discrete model is positive defined, nevertheless it has been observed that this property is not preserved in the continuum model obtained from such approach (Rosenau, 2003, Kumar and McDowell, 2004, Bacigalupo and Gambarotta, 2017b). Consequently, loss of ellipticity inducing spurious short-wavelength instabilities, imaginary frequencies in the elastic field and unbounded group velocities in the short-wave limit may result, just to mention the main drawbacks.

A dual approach is based on the so called *continualization of discrete governing equation* in which the difference of displacements of adjacent nodes in the equation of motion of the discrete model is approximated by a truncated series of a macro-displacement field (Askes *et al.*, 2002, Askes and Metrikine, 2005, Bacigalupo and Gambarotta, 2017.b) or by involving the shift and pseudo-differential operators and their power series approximations (Andrianov and Awrejcewicz, 2008, Bacigalupo and Gambarotta, 2019). This approach does not solve the pathologies in the constitutive equation of the equivalent continuum model mentioned above, being related to a non-positive defined potential energy density (Metrikine and Askes, 2002, Bacigalupo and Gambarotta, 2017b). It is worth to note that, in general, these continualization techniques lead to governing equations involving local inertial terms and non-local constitutive tensors.

*Enhanced or regularized continualization methods* have been proposed to circumvent the above-mentioned pathologies. One of the most common approaches consists in transforming the governing equations of the discrete system into pseudo-differential equations using shift operators and therefore approximating these equations through a Padé approximant. In these cases, non-local inertial terms are included in the equation of motion of the equivalent continuum, namely spatial derivatives of the continuum acceleration field appear, which are indirectly introduced by the Padé approximant applied to the constitutive terms in the continuum differential problem (Kevrekidis *et al.*, 2002, Andrianov and Awrejcewicz, 2005, 2008, Lombardo and Askes, 2010, Challamel *et al.*, 2018). However, as shown by Bacigalupo and Gambarotta (2019), in some cases this approach cannot avoid non-positive definiteness of the elastic potential energy density of the homogenized continuum. It is worth to note that the introduction of inertial non-locality involves enriched models, characterized by constitutive and inertial characteristic lengths, in the spirit of the non-local continuum models proposed by the seminal papers of Mindlin (1964), Eringen (1983), Askes and Aifantis (2011). The non-local constitutive tensors may be phenomenologically assumed or deduced in a closed form and consistent way through dynamic variational-asymptotic homogenization schemes (see Bacigalupo and Gambarotta, 2014), analyzed in detail in Rosi and Auffray (2016). A generalization of the constitutive equation proposed by Askes and Aifantis (2011) is proposed in De Domenico and Askes (2016) and De Domenico *et al.* (2019), in which the field equation governing the continuum model is recovered through a continualization technique based on the Padé approximant applied to the discrete equations of one-dimensional lattice.

A dynamic asymptotic homogenization to obtain second order continuum models with local inertial terms has been proposed by Reda *et al.* (2016) to approximate the frequency band structure of discrete elastic systems. This technique is based on asymptotic development of the kinematic variables and on the lattice equilibrium equations in virtual power form and provides approximations of the acoustic branches of the frequency spectrum. A further high contrast homogenization technique to obtain a high frequency approximation of the band structure of periodic lattice materials has been elegantly and consistently formulated in Kamotski and Smyshlyaev (2019).

Schemes based on enhanced continualization have been proposed, in which the continuous macro-displacement field is related to a finite difference of a certain order and precision $n$ expressed in terms of the nodal displacements of the Lagrangian model. Rosenau (2003) proposed a first order regularization approach applied to a monoatomic chain based on a finite forward difference of accuracy $n = 1$ in which non-local inertial terms are consistently introduced in the governing equations of the equivalent continuum. A generalized micropolar enhanced continualization of one-dimensional beam lattices has been proposed by Bacigalupo and Gambarotta (2019), where a first order regularization approach based on the central difference with accuracy $n = 2$ was used to get pseudo-differential field equations at the macroscale based on the concept of shift operators. Here, through a formal power series expansion of these equations, the governing differential equations have been derived, involving inertial and constitutive non-local terms with corresponding Lagrangian functionals energetically consistent. The continuum model thus obtained provides good approximations of the static and dynamic response of the discrete model.

Finally, other continualization techniques have been proposed based on the spatial discrete Fourier transforms or on the bilateral Z-transform of the equations of motion of one-dimensional Lagrangian systems. Several Authors have shown that this approach provides consistent integral-differential equations of the equivalent non-local continuum in terms of the continuous displacement field (see Kunin, 1982, Andrianov *et al.*, 2012, Charlotte and Truskinovsky, 2012, Bacigalupo and Gambarotta, 2019). The resulting integral-differential equation can be approximated with higher order differential equations by expanding the integral kernel in power series or through a Padè approximant. In some cases, the resulting Lagrangian functional of the continuum model turns out to have elastic potential and kinetic energy density not-positive defined.

In this context, the present formulation is focused on the dynamical identification of equivalent non-local continua representative of mono and two-dimensional lattices with lumped mass at the nodes. To this aim, the frequency band structure of the equivalent continualized system is required to coincide with the one of the effective discrete model, namely the spatial bilateral Z-transform of the nodal displacements, appropriately mapped on the unit circle, is matched to the Fourier transform of the macroscale displacement field. In the case of one-dimensional monoatomic lattice, the derived integral-differential equation turns out to coincide with the pseudo-differential equation by Bacigalupo and Gambarotta (2019). If the kernel is approximated through power series, non-local higher-order differential field equations are achieved and the constitutive part has, in general, non-positive definite potential energy. This drawback is here overcome by introducing a continuous regularized auxiliary displacement field. The integral-differential governing equations and the corresponding differential equations are restated, which are characterized by constitutive and inertial non-locality. The proposed approach is generalized in a consistent way to two-dimensional systems through the introduction of multi-dimensional bilateral Z- and Fourier transforms and a suitable generalization of the continuous regularized auxiliary field, a process that may be easily extended to three-dimensional lattices. Finally, two examples of lattice-like systems consisting of periodic pre-stressed cable-nets of mass points are analyzed. The resulting higher order continua characterized by differential equations obtained as approximations of the integral-differential ones show dispersion surfaces in very good agreement with those of the Lagrangian effective model.

## 2. Standard versus enhanced continualization of one-dimensional lattice materials

To introduce the enhanced continualization let consider two one-dimensional simple systems made up of equally spaced nodes with mass $m$ and connected through massless axial ligaments of length $\ell$. The first system is the monoatomic lattice with ligaments of axial stiffness $h$ linking the nodal masses undergoing axial displacement and subjected to nodal axial forces (Figure 1.a). The second system is the $N$ pre-stressed cable with the nodal masses undergoing transverse displacements and transverse applied forces (Figure 1.b). The dimensionless linearized equation of motion of the $i-\text{th}$ node of the two systems takes the classical form

$$\psi_{i-1} - 2\psi_i + \psi_{i+1} + f_i = I_\psi \ddot{\psi}_i, \tag{1}$$

having defined for system (a) the dimensionless axial displacement $\psi_i = \frac{u_i}{\ell}$, nodal force $f_i = \frac{\mathrm{f}_i}{h\ell}$, nodal mass $I_\psi = \frac{m}{h}$, and for system (b) the dimensionless transverse displacement $\psi_i = \frac{v_i}{\ell}$, nodal force $f_i = \frac{\mathrm{f}_i}{N}$, nodal mass $I_\psi = \frac{m\ell}{N}$. It is well known that the harmonic wave propagation along these one-dimensional systems is dispersive with angular frequency

$$\omega(k\ell) = \frac{2}{\sqrt{I_\psi}}\left|\sin\left(\frac{k\ell}{2}\right)\right| = \frac{k\ell}{\sqrt{I_\psi}}\left[1 - \frac{1}{24}(k\ell)^2 + \frac{1}{1920}(k\ell)^4\right] + \mathcal{O}\left(k^7\ell^7\right), \tag{2}$$

depending on the dimensionless wavelength $k\ell$ with periodicity $2\pi$ that represents the acoustic branch in the Bloch spectrum.

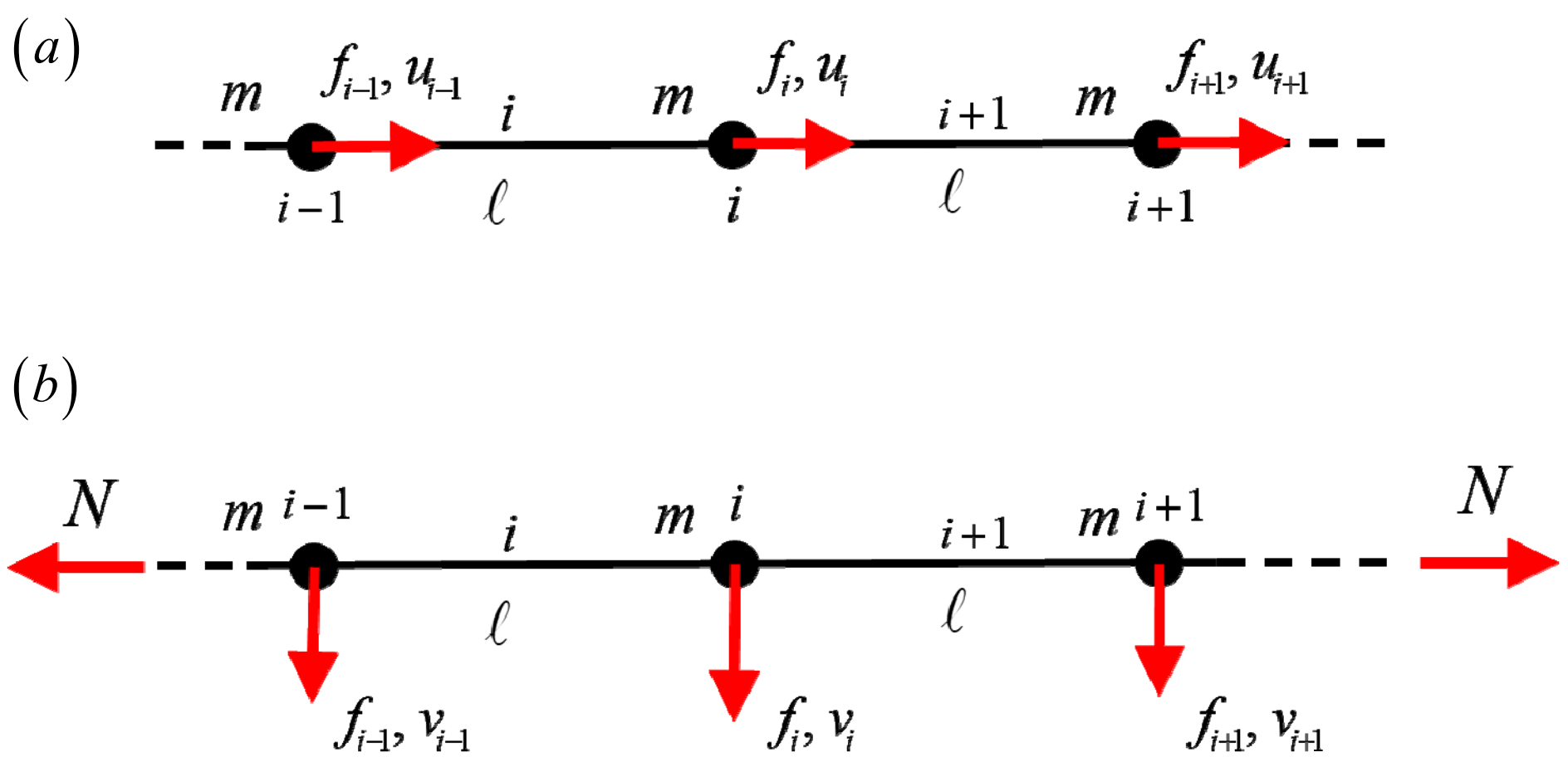

Figure 1. (a) Monoatomic lattice; (b) prestressed cable. Nodal displacements and applied forces.

Equation (1) may be reformulated by introducing the Z-transform $\mathcal{Z}[\psi_i] = \sum_{-\infty}^{+\infty} \psi_i(t) z^{-i}$, with $z \in \mathbb{C}$ (see for reference Ragazzini and Lotfi, 1956; Jury *et.al*, 1964), with the property $\mathcal{Z}[\psi_{i\pm1}] = z^{\pm1}\mathcal{Z}[\psi_i]$, as follows

$$z^{-1}\mathcal{Z}[\psi_i] - 2\mathcal{Z}[\psi_i] + z\,\mathcal{Z}[\psi_i] + \mathcal{Z}[f_i] = I_\psi \mathcal{Z}[\ddot{\psi}_i] \tag{3}$$

in the $Z$ space. The notation is simplified by defining $\mathcal{Z}[\psi_i] = \hat{\psi}(z,t)$ so that equation (3) takes the form

$$\left(z^{-1} - 2 + z\right)\hat{\psi}(z,t) + \hat{f}(z,t) = I_\psi \ddot{\hat{\psi}}(z,t)\ . \tag{4}$$

By introducing the mapping $z = \exp(Ik\ell)$ on the unit circle, with $I^2 = -1$ and $k \in \mathbb{R}$, equation (4) may be rewritten as

$$\left[\exp(-Ik\ell) - 2 + \exp(Ik\ell)\right]\hat{\psi}(k,t) + \hat{f}(k,t) = I_\psi \ddot{\hat{\psi}}(k,t) \tag{5}$$

in the Fourier space.

Let now introduce a continuum field $\Psi(x,t)$ in the physical space ($x$ being the spatial coordinate) whose Fourier transform $\mathcal{F}\left[\Psi(x,t)\right] = \int_{-\infty}^{+\infty} \Psi(x,t)\exp(-Ikx)\,dx = \hat{\Psi}(k,t)$ is assumed to coincide with the $Z$-transform of the Lagrangian displacement $\psi_i(t)$ mapped on the unit circle, namely

$$\hat{\psi}(k,t) = \mathcal{Z}[\psi_i]\big|_{z=\exp(Ik\ell)} \doteq \mathcal{F}\left[\Psi(x,t)\right] = \hat{\Psi}(k,t). \tag{6}$$

Accordingly, equation (5) may be written as

$$\left[\exp(-Ik\ell) - 2 + \exp(Ik\ell)\right]\hat{\Psi}(k,t) + \hat{F}(k,t) = I_\psi \ddot{\hat{\Psi}}(k,t) \quad . \tag{7}$$

Here, $\hat{F}(k,t)$ is the Fourier transform of the continuum field $F(x,t)$ satisfying the following property

$$\hat{F}(k,t) = \mathcal{F}\left[F(x,t)\right] \doteq \mathcal{Z}[f_i]\big|_{z=\exp(Ik\ell)} = \hat{f}(k,t), \tag{8}$$

and from the inverse Fourier transform $\mathcal{F}^{-1}\left[\hat{F}(k,t)\right] = \frac{1}{2\pi}\int_{-\infty}^{+\infty} \hat{F}(k,t) \exp(Ikx)\,\ell dk$ one obtains

$$\begin{aligned} F(x,t) &= \frac{1}{2\pi}\int_{-\infty}^{+\infty} \mathcal{F}\left[F(x,t)\right] \exp(Ikx)\,\ell dk = \\ &= \frac{1}{2\pi}\int_{-\infty}^{+\infty} \mathcal{Z}[f_i]\big|_{z=\exp(Ik\ell)} \exp(Ikx)\,\ell dk = \frac{1}{2\pi}\sum_{-\infty}^{+\infty} f_i(t)\int_{-\infty}^{+\infty} \exp\left[Ik(x-i\ell)\right]\ell dk. \end{aligned} \tag{9}$$

The inverse Fourier transform of equation (7) provides the following integral-differential equation

$$\mathcal{F}^{-1}\left[\left(\exp(-Ik\ell) - 2 + \exp(Ik\ell)\right)\hat{\Psi}(k,t)\right] + \mathcal{F}^{-1}\left[\hat{F}(k,t)\right] = I_\psi \mathcal{F}^{-1}\left[\ddot{\hat{\Psi}}(k,t)\right], \tag{10}$$

that is specialized as

$$\begin{aligned} &\frac{1}{2\pi}\int_{-\infty}^{+\infty}\left[\exp(-Ik\ell) - 2 + \exp(Ik\ell)\right]\mathcal{F}\left[\Psi(x,t)\right] \exp(Ikx)\,\ell dk + F(x,t) = \\ &\quad = I_\psi \frac{1}{2\pi}\int_{-\infty}^{+\infty} \mathcal{F}\left[\ddot{\Psi}(x,t)\right] \exp(Ikx)\,\ell dk. \end{aligned} \tag{11}$$

In order to obtain an higher order differential governing equation the kernel $\mathcal{K}_1(k\ell) = \exp(-Ik\ell) - 2 + \exp(Ik\ell) = 2\cos(k\ell) - 2$ may be expressed through a Taylor series and equation (11) can be rewritten in the form

$$\begin{aligned} &\frac{1}{2\pi}\int_{-\infty}^{+\infty}\left[-2 + 2\sum_{n=0}^{+\infty}\frac{(Ik\ell)^{2n}}{(2n)!}\right]\mathcal{F}\left[\Psi(x,t)\right] \exp(Ikx)\,\ell dk + F(x,t) = \\ &\quad = I_\psi \frac{1}{2\pi}\int_{-\infty}^{+\infty} \mathcal{F}\left[\ddot{\Psi}(x,t)\right] \exp(Ikx)\,\ell dk. \end{aligned} \tag{12}$$

By noting that $\mathcal{K}_1(k\ell)$ is an entire analytic function (defined and analytic on the entire space of the dimensionless wave number $k\ell$, considered as complex variable), the radius of convergence of its Taylor series is infinite and it is possible to reverse the order of integration and summation. Moreover, by retaining the first $M$ terms of the series and after a simple algebra one gets

$$\begin{aligned}&\frac{1}{2\pi}\sum_{n=1}^{M}\int_{-\infty}^{+\infty}\left[\frac{2(Ik\ell)^{2n}}{(2n)!}\right]\mathcal{F}\left[\Psi(x,t)\right]\exp(Ikx)\,\ell dk+F(x,t)=\\&=I_\psi\frac{1}{2\pi}\int_{-\infty}^{+\infty}\mathcal{F}\left[\ddot{\Psi}(x,t)\right]\exp(Ikx)\,\ell dk.\end{aligned}\tag{13}$$

Recalling the property of the inverse Fourier transform of a generic function $f(x)$

$$\mathcal{F}^{-1}\left[(Ik)^m\mathcal{F}\left[f(x)\right]\right]=\mathcal{F}^{-1}\left[(Ik)^m\hat{f}(k)\right]=\frac{d^m f(x)}{dx^m},\qquad m\in\mathbb{N}^*,\tag{14}$$

equation (13) is specialized in the form

$$\sum_{n=1}^{M}\frac{2\ell^{2n}}{(2n)!}\frac{\partial^{2n}\Psi(x,t)}{\partial x^{2n}}+F(x,t)=I_\psi\ddot{\Psi}(x,t),\tag{15}$$

which is the governing equation of motion of the continuum of order $M$.

The same result may be obtained by recalling the notion of pseudo-differential operators (see for reference Maslov, 1976, Shubin, 1987, Bacigalupo and Gambarotta, 2019) allowing the integral-differential equation (11) written as a pseudo-differential equation in the form

$$\left[\exp(-\ell D)-2+\exp(\ell D)\right]\Psi(x,t)+F(x,t)=I_\psi\ddot{\Psi}(x,t),\tag{16}$$

where the differential operator $D(\cdot)=\frac{\partial}{\partial x}(\cdot)$ is introduced. Equation (16) may be obtained through a different, but equivalent, approach based on the concept of shift $E_\ell$ linking the displacement of two adjacent nodes $\psi_{i+1}=E_\ell\ \psi_i$ (see for reference Jordan, 1965, Rota *et al.*, 1973, Kelley and Peterson, 2001, Bacigalupo and Gambarotta, 2019). By expanding the pseudo-differential operator $P(D)=\left[\exp(-\ell D)-2+\exp(\ell D)\right]$ in a formal Taylor series in the $D$ variable and by retaining its first $M$ terms, the higher order differential equation (15) of the equivalent continuum are obtained, which correspond to those derived via *standard* *continualization* (see Bacigalupo and Gambarotta, 2019). While these higher order models provide a rather accurate description of the dispersion function in the long-wave limit, on the other side they present the drawback to be the Euler-Lagrange equation associated to a Lagrangian functional having energy potential density not positive defined as a consequence of the positive values of the coefficients $2\ell^{2n}/(2n)!$ with $n \in \mathbb{N}^*$ in equation (15).

Such kind of problems may be circumvented by introducing a proper defined mapping relating the transformed displacements $\hat{\Psi}(k,t)$ in the Fourier space with a new auxiliary regularizing continuum field $\hat{\Psi}^R(k,t)$ in the same space as follows

$$\hat{\Psi}^R(k,t) \doteq \frac{\exp(Ik\ell) - \exp(-Ik\ell)}{2Ik\ell}\hat{\Psi}(k,t). \tag{17}$$

It may be shown that the above assumption has an equivalent one in the physical space by relating the first derivative the continuum regularized field $\Psi^R(x,t) = \mathcal{F}^{-1}\left[\hat{\Psi}^R(k,t)\right]$ to the central difference of the discrete variables $\left.\frac{\partial \Psi^R}{\partial x}\right|_{x_i} \doteq \frac{\psi_{i+1} - \psi_{i-1}}{2\ell}$ as shown in Appendix A.

From equation (17), the equation of motion in the Fourier transformed space (7) may be rewritten as

$$\frac{2Ik\ell\left[\exp(-Ik\ell) - 2 + \exp(Ik\ell)\right]}{\exp(Ik\ell) - \exp(-Ik\ell)}\hat{\Psi}^R(k,t) + \hat{F}(k,t) = I_\psi \frac{2Ik\ell}{\exp(Ik\ell) - \exp(-Ik\ell)}\ddot{\hat{\Psi}}^R(k,t). \tag{18}$$

The inverse Fourier transform of equation (18) provides the following integral-differential equation

$$\begin{aligned} &\mathcal{F}^{-1}\left[\frac{2Ik\ell\left[\exp(-Ik\ell) - 2 + \exp(Ik\ell)\right]}{\exp(Ik\ell) - \exp(-Ik\ell)}\mathcal{F}\left[\Psi^R(x,t)\right]\right] + \mathcal{F}^{-1}\left[\hat{F}(k,t)\right] = \\ &= I_\psi \mathcal{F}^{-1}\left[\frac{2Ik\ell}{\exp(Ik\ell) - \exp(-Ik\ell)}\mathcal{F}\left[\ddot{\Psi}^R(x,t)\right]\right], \end{aligned} \tag{19}$$

that is specialized as

$$\begin{aligned} &\frac{1}{2\pi}\int_{-\infty}^{+\infty}\frac{2Ik\ell\left[\exp(-Ik\ell) - 2 + \exp(Ik\ell)\right]}{\exp(Ik\ell) - \exp(-Ik\ell)}\mathcal{F}\left[\Psi^R(x,t)\right]\exp(Ikx)\ell dk + F(x,t) = \\ &= I_\psi \frac{1}{2\pi}\int_{-\infty}^{+\infty}\frac{2Ik\ell}{\exp(Ik\ell) - \exp(-Ik\ell)}\mathcal{F}\left[\ddot{\Psi}^R(x,t)\right]\exp(Ikx)\ell dk. \end{aligned} \tag{20}$$

In order to obtain a generalized higher order governing equation, the numerator and the denominator of the constitutive kernel $\mathcal{K}_2(k\ell)=\dfrac{2Ik\ell\left[\exp(-Ik\ell)-2+\exp(Ik\ell)\right]}{\exp(Ik\ell)-\exp(-Ik\ell)}=\dfrac{2k\ell\left[\cos(k\ell)-1\right]}{\sin(k\ell)}$ and the inertial kernel $\mathcal{K}_3(k\ell)=\dfrac{2Ik\ell}{\exp(Ik\ell)-\exp(-Ik\ell)}=\dfrac{k\ell}{\sin(k\ell)}$, are expanded in Taylor series and the equation (20) is rewritten as follows

$$\begin{aligned}&\frac{1}{2\pi}\int_{-\infty}^{+\infty}2\left[-\frac{Ik\ell}{\sum_{n=0}^{+\infty}\frac{(Ik\ell)^{2n+1}}{(2n+1)!}}+\frac{\sum_{n=0}^{+\infty}\frac{(Ik\ell)^{2n+1}}{(2n)!}}{\sum_{n=0}^{+\infty}\frac{(Ik\ell)^{2n+1}}{(2n+1)!}}\right]\mathcal{F}\left[\Psi^R(x,t)\right]\exp(Ikx)\ell dk+F(x,t)=\\&=I_\psi\frac{1}{2\pi}\int_{-\infty}^{+\infty}\left[\frac{Ik\ell}{\sum_{n=0}^{+\infty}\frac{(Ik\ell)^{2n+1}}{(2n+1)!}}\right]\mathcal{F}\left[\ddot{\Psi}^R(x,t)\right]\exp(Ikx)\ell dk.\end{aligned}\tag{21}$$

By remembering the concept of discrete convolution of two series and some algebra, the governing equation takes the following form

$$\begin{aligned}&\frac{1}{2\pi}\int_{-\infty}^{+\infty}\left[\sum_{n=0}^{+\infty}(c_n-2e_n)(Ik\ell)^{2n}\right]\mathcal{F}\left[\Psi^R(x,t)\right]\exp(Ikx)\ell dk+F(x,t)=\\&=I_\psi\frac{1}{2\pi}\int_{-\infty}^{+\infty}\left[\sum_{n=0}^{+\infty}e_n(Ik\ell)^{2n}\right]\mathcal{F}\left[\ddot{\Psi}^R(x,t)\right]\exp(Ikx)\ell dk,\end{aligned}\tag{22}$$

being the terms $c_n$ and $e_n$ given by the recursive relations $c_n=\dfrac{1}{a_0}\left[b_n-\sum_{j=0}^{n-1}a_{n-j}c_j\right]$ and $e_n=-\dfrac{1}{a_0}\sum_{j=0}^{n-1}a_{n-j}e_j$, where $a_n=1/(2n+1)!$ and $b_n=2/(2n)!$ with $n\in\mathbb{N}^*$. The region of convergence of the power series of the kernels $\mathcal{K}_2(k\ell)$ and $\mathcal{K}_3(k\ell)$ about $k\ell=0$ is shown as a red circle in Figure 2.a. and 2.b, respectively, in the space of the dimensionless wave number $k\ell$, here considered as complex variable. More specifically, the radius of convergence of both the power series is proved to be $\pi$. Indeed, the kernels $\mathcal{K}_2(k\ell)$, $\mathcal{K}_3(k\ell)$ have poles periodically distributed along the real axis $\Im m(k\ell)=0$, and the singularities closest to $k\ell=0$ are found in $k\ell=\pm\pi$ which demarcate the limit of the first Brillouin zone.

In Figure 2a and 2b the magnitudes $|\mathcal{K}_2| = \sqrt{(\Re e(\mathcal{K}_2))^2 + (\Im m(\mathcal{K}_2))^2}$, $|\mathcal{K}_3| = \sqrt{(\Re e(\mathcal{K}_3))^2 + (\Im m(\mathcal{K}_3))^2}$ are shown in terms of $\Re e(k\ell)$, $\Im m(k\ell)$ and the poles in the range $-3\pi \le \Re e(k\ell) \le 3\pi$ are localized at the blue points. The real and imaginary parts of the kernels $\mathcal{K}_2(k\ell)$, $\mathcal{K}_3(k\ell)$ with their magnitudes are given in Appendix B as the dimensionless complex wave number $k\ell = \Re e(k\ell) + I\Im m(k\ell)$ varies. Moreover, in Figure 2c and 2d the real part of $\mathcal{K}_2(k\ell)$ and $\mathcal{K}_3(k\ell)$ are plotted in terms of $\Re e(k\ell)$ where one may observe the approximations of the kernels in power series around $k\ell = 0$. As expected, increasing the order of the power series the convergence to the functions $\Re e(\mathcal{K}_2)$ and $\Re e(\mathcal{K}_3)$ is observed in the first Brillouin zone. Accordingly, inside the red circle it is possible to reverse the order of integration and summation in equation (22). By retaining the first $M$ terms in the series one gets after simple algebra

$$\begin{aligned} &\frac{1}{2\pi}\sum_{n=1}^{M}(c_n - 2e_n)\int_{-\infty}^{+\infty}(Ik\ell)^{2n}\,\mathcal{F}\left[\Psi^R(x,t)\right]\exp(Ikx)\,\ell dk + F(x,t) = \\ &= I_\psi \frac{1}{2\pi}\sum_{n=0}^{M} e_n \int_{-\infty}^{+\infty}(Ik\ell)^{2n}\,\mathcal{F}\left[\ddot{\Psi}^R(x,t)\right]\exp(Ikx)\,\ell dk \;, \end{aligned} \tag{23}$$

where the identity $c_0 - 2e_0 = 0$ has been applied. According to property (14) the governing equation (23) may be specialized in the form

$$\sum_{n=1}^{M}\ell^{2n}(c_n - 2e_n)\,\frac{\partial^{2n}\Psi^R(x,t)}{\partial x^{2n}} + F(x,t) = I_\psi \sum_{n=0}^{M}\ell^{2n}\, e_n \frac{\partial^{2n}\ddot{\Psi}^R(x,t)}{\partial x^{2n}}, \tag{24}$$

which is the governing equation of motion of the continuum of order $M$ characterized by non-local constitutive and inertial terms. The coefficients of the power series for $n \le 6$ are given in Table I, where it may be observed that such coefficients are characterized by alternate sign with $n$, so implying a positive definite kinetic and potential energy density.

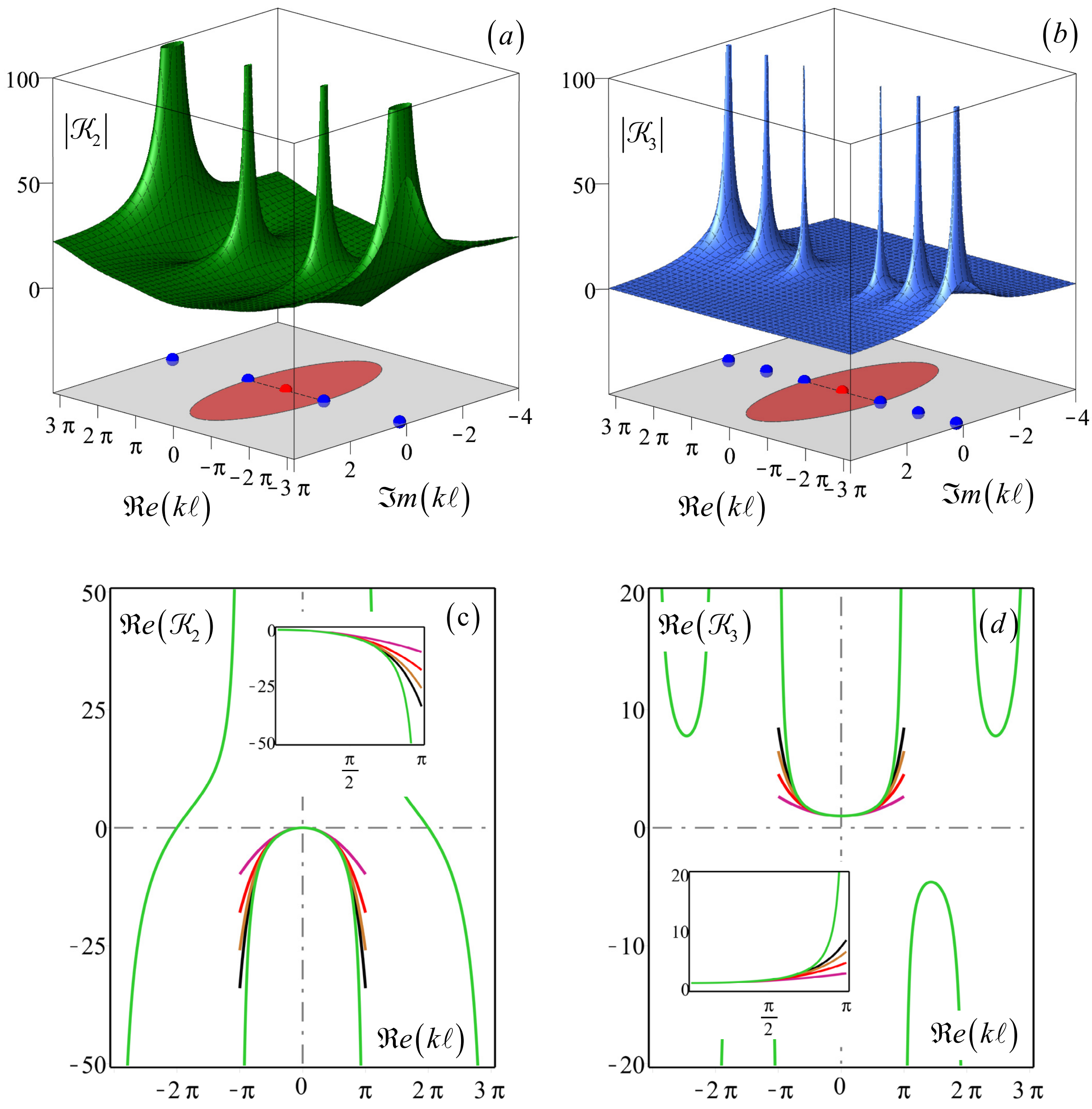

Figure 2 (a) Magnitude of the constitutive kernel $\mathcal{K}_2(k\ell)$; (b) magnitude of the inertial kernel $\mathcal{K}_3(k\ell)$; (c), (d) real parts of the kernels $\mathcal{K}_2(k\ell)$, $\mathcal{K}_3(k\ell)$ (green line) and their approximations via Taylor polynomial around $k\ell = 0$: 2$^{nd}$ order (violet line), 4$^{th}$ order (red line), 6$^{th}$ order (yellow line), 8$^{th}$ order (black line).

Table I. Coefficients of the power series of the governing equation (24).

| $n$ | $e_n$ | $c_n - 2e_n$ |
|---|---|---|
| 0 | $1$ | $0$ |
| 1 | $-\frac{1}{6}$ | $1$ |
| 2 | $\frac{7}{360}$ | $-\frac{1}{12}$ |
| 3 | $-\frac{31}{15120}$ | $\frac{1}{120}$ |
| 4 | $\frac{127}{604800}$ | $-\frac{17}{20160}$ |
| 5 | $-\frac{73}{3421440}$ | $\frac{31}{362880}$ |
| 6 | $\frac{1414477}{653837184000}$ | $-\frac{691}{79833600}$ |

As previously shown, the same result may be obtained if the concept of pseudo-differential operator is introduced in equation (20) which takes the equivalent form

$$\begin{aligned}&\left[\frac{2\ell D\left(\exp(-\ell D)-2+\exp(\ell D)\right)}{\exp(\ell D)-\exp(-\ell D)}\right]\Psi^R(x,t)+F(x,t)=\\&=I_\psi\left[\frac{2\ell D}{\exp(\ell D)-\exp(-\ell D)}\right]\ddot{\Psi}^R(x,t),\end{aligned}\tag{25}$$

involving two pseudo-differential operators $P_1(D)=\frac{2\left[\exp(\ell D)-2+\exp(-\ell D)\right]}{\left[\exp(\ell D)-\exp(-\ell D)\right]}\ell D$ and $P_2(D)=\frac{2\ell D}{\left[\exp(\ell D)-\exp(-\ell D)\right]}$, that coincides with the governing equation obtained via *enhanced continualization* with first order regularization approach proposed by Bacigalupo and Gambarotta (2019). Higher order differential field equations may be obtained through Taylor expansions in $k$ variable of the kernels of the integral-differential equation (20) or, alternatively, through a formal Taylor expansion in $D$ variable of the pseudo-differential, the latter being written in the following form

$$P_1(D) = \ell^2 D^2 - \frac{1}{12}\ell^4 D^4 + \frac{1}{120}\ell^6 D^6 + \mathcal{O}\left(\ell^7 D^7\right),$$
$$P_2(D) = 1 - \frac{1}{6}\ell^2 D^2 + \frac{7}{360}\ell^4 D^4 - \frac{31}{15120}\ell^6 D^6 + \mathcal{O}\left(\ell^7 D^7\right), \tag{26}$$

with the coefficients coincident with those ones in Table I.

A first approximation of the equation of motion of the equivalent continuum is obtained from equation (25) once the terms up to the second order in (26) are retained

$$\ell^2 \frac{\partial^2 \Psi^R}{\partial x^2} + F(x,t) = I_\psi \left( \ddot{\Psi}^R - \frac{\ell^2}{6} \frac{\partial^2 \ddot{\Psi}^R}{\partial x^2} \right). \tag{27}$$

It is worth to note that equation (27) differs from the corresponding one from the standard continualization in the non-local inertia term. With respect to the equation of motion derived by applying the Pade's approximant (see Cuyt, 1980; Kevrekidis et al., 2002; De Domenico and Askes, 2018) the difference is found in the coefficient of the non-local inertia term. The dispersion function deriving from equation (27) is

$$\omega = \frac{k\ell}{\sqrt{I_\psi}} \sqrt{\frac{1}{\left(1 + \frac{1}{6}k^2\ell^2\right)}} = \frac{k\ell}{\sqrt{I_\psi}} \left( 1 - \frac{1}{12}k^2\ell^2 + \frac{1}{96}k^4\ell^4 \right) + \mathcal{O}\left(k^5\ell^5\right). \tag{28}$$

If terms up to the fourth order are retained in (26), the continualized equation of motion takes the form

$$\ell^2 \frac{\partial^2 \Psi^R}{\partial x^2} - \frac{\ell^4}{12} \frac{\partial^4 \Psi^R}{\partial x^4} + F(x,t) = I_\psi \left( \ddot{\Psi}^R - \frac{\ell^2}{6} \frac{\partial^2 \ddot{\Psi}^R}{\partial x^2} + \frac{7\ell^4}{360} \frac{\partial^4 \ddot{\Psi}^R}{\partial x^4} \right), \tag{29}$$

and the resulting dispersion function is

$$\omega = \frac{k\ell}{\sqrt{I_\psi}} \sqrt{\frac{1 + \frac{1}{12}k^2\ell^2}{1 + \frac{k^2\ell^2}{6} + \frac{7}{360}k^4\ell^4}} = \frac{k\ell}{\sqrt{I_\psi}} \left( 1 - \frac{1}{24}k^2\ell^2 - \frac{7}{1920}k^4\ell^4 \right) + \mathcal{O}\left(k^5\ell^5\right). \tag{30}$$

It easy to show that both the potential energy density and the kinetic energy density of the Lagrangian associated to the Euler-Lagrange equation are positive defined independently of the truncation order of the expansion of the pseudo-operators (26). As an example, see the Lagrangian density function

$$\mathcal{L}=\frac{1}{2\ell}I_{\psi}\left[\left(\dot{\Psi}^{R}\right)^{2}+\frac{\ell^{2}}{6}\left(\frac{\partial\dot{\Psi}^{R}}{\partial x}\right)^{2}+\frac{7\ell^{4}}{360}\left(\frac{\partial^{2}\dot{\Psi}^{R}}{\partial x^{2}}\right)^{2}\right]+$$
$$-\frac{1}{2\ell}\left[\ell^{2}\left(\frac{\partial\Psi^{R}}{\partial x}\right)^{2}+\frac{\ell^{4}}{12}\left(\frac{\partial^{2}\Psi^{R}}{\partial x^{2}}\right)^{2}\right]+\frac{1}{\ell}F(x,t)\Psi^{R} \quad , \tag{31}$$

associated to the equation of motion (29) having positive defined the potential and the kinetic energy density.

The Floquet-Bloch spectra of the continuum model obtained via enhanced continualization have been compared with good agreement with those by the Lagrangian model in the Figure 2 of Bacigalupo and Gambarotta (2019). The convergence of the dispersive functions obtained by the enhanced continualization to the exact one is here theoretically confirmed based on the analysis of the radius of convergence of the power series representative of the constitutive and the inertial kernels (see Figure 2).

## 3. Extension to 2D lattice materials

The continualization procedure presented in the previous Section is here extended to two-dimensional lattices. To get a simplified treatment without loss of generality, let us consider the transverse vibrations of a periodic prestressed cable-net of mass points obtained by the repetition of centro-symmetric periodic cells along given periodicity vectors (see for instance Rosenau, 1987, Andrianov and Awrejewicz, 2008). Each cell contains a node which is connected to the nodes of the $n$ adjacent cells through equally prestressed strings of length $\ell$, $n$ being the coordination number of the lattice. Let us consider the reference node $\left(i_1,...,i_{n/2}\right)$ and two opposite nodes with respect to it connected by two opposite strings along unit vector $\mathbf{n}^{(j)}$, with $j=1,..,n/2$; these opposite nodes are denoted by $\left(i_1,..,i_j+1,..,i_{n/2}\right)$ and $\left(i_1,..,i_j-1,..,i_{n/2}\right)$, respectively (see Figure 3). It is worth noting that the indices that identify the nodes represents the directions of coordination of the lattice starting from the reference node. Let us denote with $\psi_{i_1,...,i_{n/2}}$ the dimensionless nodal displacement of the reference node and with $\psi_{i_1,..,i_j+1,..,i_{n/2}}$ and $\psi_{i_1,..,i_j-1,..,i_{n/2}}$ the displacements of nodes $\left(i_1,..,i_j+1,..,i_{n/2}\right)$ and $\left(i_1,..,i_j-1,..,i_{n/2}\right)$.

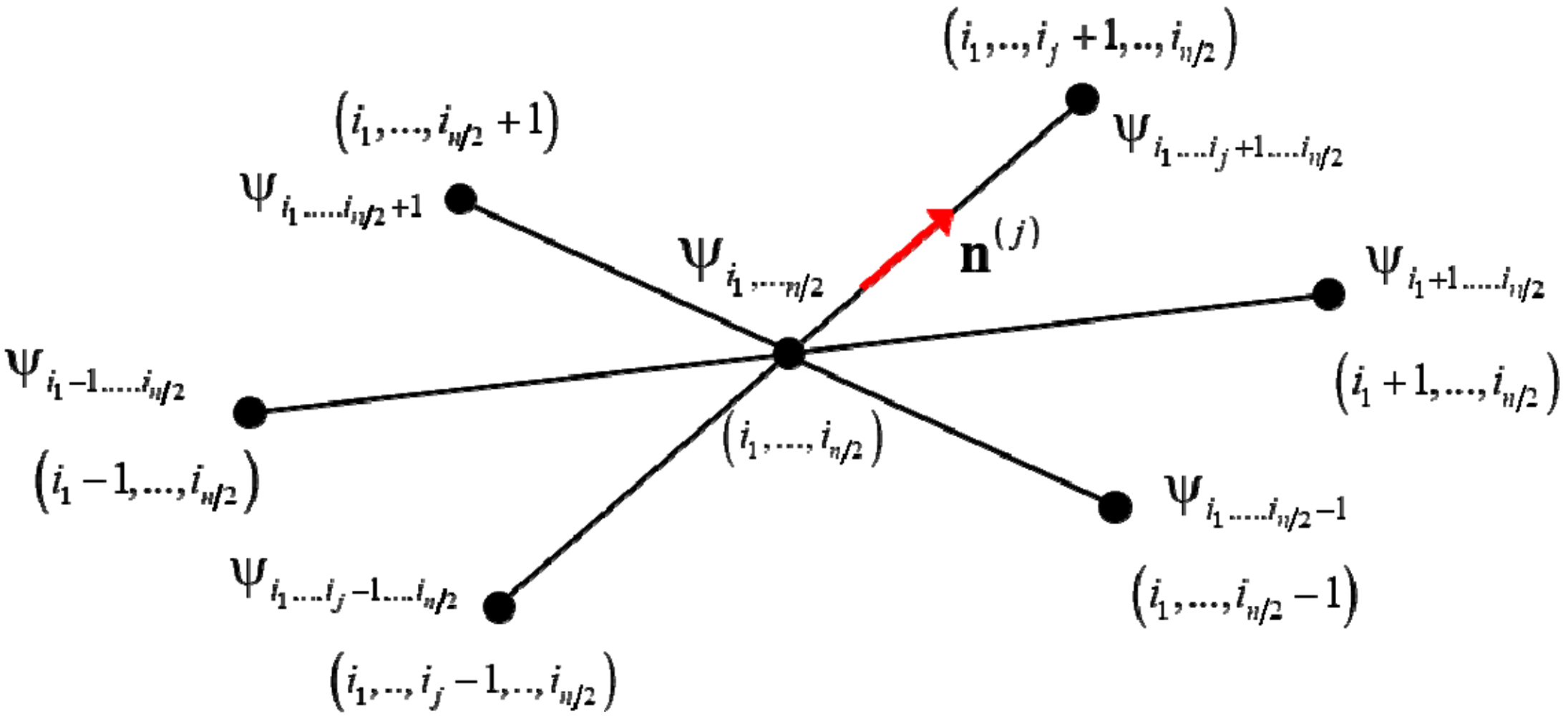

Figure 3. The reference node and the adjacent ones together with the corresponding dimensionless transverse displacements.

The linearized equation of the transverse motion of the reference node in the Lagrangian system may be written as a linear difference equation in the form

$$\sum_{\mathscr{P}(i_1,\dots,i_{n/2})} \alpha_{i_1,\dots,i_{n/2}} \psi_{i_1,\dots,i_{n/2}} + f_{i_1,\dots,i_{n/2}} = I_\psi \ddot{\psi}_{i_1,\dots,i_{n/2}}, \tag{32}$$

where the symbol $\mathscr{P}(i_1,\dots,i_{n/2})$ denotes the set of the characteristic nodes of the lattice, namely the reference node and the adjacent ones, $\alpha_{i_1,\dots,i_{n/2}}$ the coefficient depending on the network topology, $f_{i_1,\dots,i_{n/2}}$ is related to the applied force to the reference node and $I_\psi = \dfrac{m\ell}{N}$ represents the nodal inertia, in agreement with the notation introduced in Section 2. Specifically, the equation of motion may be written in an extended form

$$\begin{aligned} &\alpha_{i_1-1,\dots,i_{n/2}} \psi_{i_1-1,\dots,i_{n/2}} + \alpha_{i_1+1,\dots,i_{n/2}} \psi_{i_1+1,\dots,i_{n/2}} + \alpha_{i_1,\dots,i_{n/2}} \psi_{i_1,\dots,i_{n/2}} + \dots\dots + \\ &+\alpha_{i_1,\dots,i_{n/2}-1} \psi_{i_1,\dots,i_{n/2}-1} + \alpha_{i_1,\dots,i_{n/2}+1} \psi_{i_1,\dots,i_{n/2}+1} + f_{i_1,\dots,i_{n/2}} = I_\psi \ddot{\psi}_{i_1,\dots,i_{n/2}}. \end{aligned} \tag{33}$$

Following the approach in Section 2, the multi-dimensional bilateral Z-transform $\mathscr{Z}\left[\psi_{i_1,\dots,i_{n/2}}\right] = \sum_{-\infty}^{+\infty} \cdots \sum_{-\infty}^{+\infty} \psi_{i_1,\dots,i_{n/2}}(t) z_1^{-i_1} \cdots z_{n/2}^{-i_{n/2}}$ may be introduced, with $z_j \in \mathbb{C}$ , having the property $\mathscr{Z}\left[\psi_{i_1 \pm m_1,\dots,i_{n/2} \pm m_{n/2}}\right] = z_1^{\pm m_1} \cdots z_{n/2}^{\pm m_{n/2}} \mathscr{Z}\left[\psi_{i_1,\dots,i_{n/2}}\right]$ (see for reference Tzafestas, 1986, Gregor, 1988). Accordingly, equation (33) may be reformulated as follows

$$\mathcal{Z}\left[\sum_{\mathcal{P}(i_1,\dots,i_{n/2})}\alpha_{i_1,\dots,i_{n/2}}\psi_{i_1,\dots,i_{n/2}}\right]+\mathcal{Z}\left[f_{i_1,\dots,i_{n/2}}\right]=I_\psi\mathcal{Z}\left[\ddot{\psi}_{i_1,\dots,i_{n/2}}\right] \tag{34}$$

in the $Z$ space. The notation is simplified by defining $\mathcal{Z}\left[\psi_{i_1,\dots,i_{n/2}}\right]=\hat{\psi}\left(z_1,..,z_{n/2},t\right)$ and noting that

$$\begin{gathered}\mathcal{Z}\left[\sum_{\mathcal{P}(i_1,\dots,i_{n/2})}\alpha_{i_1,\dots,i_{n/2}}\psi_{i_1,\dots,i_{n/2}}\right]= \\ =\left(\alpha_{i_1-1,\dots,i_{n/2}}z_1^{-1}+\alpha_{i_1+1,\dots,i_{n/2}}z_1+\alpha_{i_1,\dots,i_{n/2}}+.......+\alpha_{i_1,\dots,i_{n/2}-1}z_{n/2}^{-1}+\alpha_{i_1,\dots,i_{n/2}+1}z_{n/2}\right)\hat{\psi}\left(z_1,..,z_{n/2},t\right)\end{gathered} \tag{35}$$

so that equation (34) takes the form in the Z-space

$$\begin{gathered}\left(\alpha_{i_1-1,\dots,i_{n/2}}z_1^{-1}+\alpha_{i_1+1,\dots,i_{n/2}}z_1+\alpha_{i_1,\dots,i_{n/2}}+.......+\alpha_{i_1,\dots,i_{n/2}-1}z_{n/2}^{-1}+\alpha_{i_1,\dots,i_{n/2}+1}z_{n/2}\right)\hat{\psi}\left(z_1,..,z_{n/2},t\right)+ \\ +\hat{f}\left(z_1,..,z_{n/2},t\right)=I\ddot{\hat{\psi}}\left(z_1,..,z_{n/2},t\right).\end{gathered} \tag{36}$$

By introducing the mapping $z_j=\exp\left(I\ n_r^{(j)}k_r\ell\right)$ on the unit circle, with $j=1,\dots,n/2$ and $r=1,2$, the wave vector component $k_r\in\mathbb{R}$ is introduced and equation (36) may be rewritten as

$$\begin{gathered}\begin{pmatrix}\alpha_{i_1-1,\dots,i_{n/2}}\exp\left(-I\ n_r^{(1)}k_r\ell\right)+\alpha_{i_1+1,\dots,i_{n/2}}\exp\left(I\ n_r^{(1)}k_r\ell\right)+\alpha_{i_1,\dots,i_{n/2}}+.......+ \\ +\alpha_{i_1,\dots,i_{n/2}-1}\exp\left(-I\ n_r^{(n/2)}k_r\ell\right)+\alpha_{i_1,\dots,i_{n/2}+1}\exp\left(I\ n_r^{(n/2)}k_r\ell\right)\end{pmatrix}\hat{\psi}\left(k_1,k_2,t\right)+ \\ +\hat{f}\left(k_1,k_2,t\right)=I\ddot{\hat{\psi}}\left(k_1,k_2,t\right)\end{gathered} \tag{37}$$

in the Fourier space.

Let now introduce a continuum field $\Psi\left(x_1,x_2,t\right)$ in the physical space (here the position vector $\mathbf{x}=\left\{x_1\quad x_2\right\}^T$ is denoted) whose Fourier transform $\mathcal{F}\left[\Psi\left(x_1,x_2,t\right)\right]=\int_{-\infty}^{+\infty}\int_{-\infty}^{+\infty}\Psi\left(x_1,x_2,t\right)\exp\left(-Ik_rx_r\right)d\mathbf{x}=\hat{\Psi}\left(k_1,k_2,t\right)$ is assumed to coincide with the $Z$-transform of the Lagrangian displacement $\psi_{i_1,\dots,i_{n/2}}\left(t\right)$ mapped on the unit circle, namely

$$\hat{\psi}\left(k_1,k_2,t\right)=\mathcal{Z}\left[\psi_{i_1,\dots,i_{n/2}}\right]\Big|_{z_j=\exp\left(In_r^{(j)}k_r\ell\right)}\doteq\mathcal{F}\left[\Psi\left(x_1,x_2,t\right)\right]=\hat{\Psi}\left(k_1,k_2,t\right). \tag{38}$$

It follows that equation (37) may be written as

$$
\begin{pmatrix} \alpha_{i_1-1,\ldots,i_{n/2}} \exp\left(-I\ n_r^{(1)} k_r \ell\right) + \alpha_{i_1+1,\ldots,i_{n/2}} \exp\left(I\ n_r^{(1)} k_r \ell\right) + \alpha_{i_1,\ldots,i_{n/2}} + \ldots\ldots + \\ +\alpha_{i_1,\ldots,i_{n/2}-1} \exp\left(-I\ n_r^{(n/2)} k_r \ell\right) + \alpha_{i_1,\ldots,i_{n/2}+1} \exp\left(I\ n_r^{(n/2)} k_r \ell\right) \end{pmatrix} \hat{\Psi}\left(k_1,k_2,t\right) \\ + \hat{F}\left(k_1,k_2,t\right) = I_\Psi \ddot{\hat{\Psi}}\left(k_1,k_2,t\right) \quad . \tag{39}
$$

In equation (39) $\hat{F}\left(k_1,k_2,t\right)$ is the Fourier transform of the continuum field $F\left(x_1,x_2,t\right)$ satisfying the following property

$$
\hat{F}\left(k_1,k_2,t\right) = \mathcal{F}\left[F\left(x_1,x_2,t\right)\right] \doteq \hat{f}\left(k_1,k_2,t\right) = \mathcal{Z}\left[f_{i_1,\ldots,i_{n/2}}\right]\Big|_{z_j=\exp\left(I\ n_r^{(j)} k_r \ell\right)} . \tag{40}
$$

By carrying out the inverse Fourier transform $\mathcal{F}^{-1}\left[\hat{F}\left(k_1,k_2,t\right)\right] = \frac{1}{\left(2\pi\right)^2}\int_{-\infty}^{+\infty}\int_{-\infty}^{+\infty} \hat{F}\left(k_1,k_2,t\right) \exp\left(Ik_r x_r\right)\ell^2 d\mathbf{k}$ one obtains

$$
\begin{aligned}
F\left(x_1,x_2,t\right) &= \frac{1}{\left(2\pi\right)^2}\int_{-\infty}^{+\infty}\int_{-\infty}^{+\infty} \mathcal{F}\left[F\left(x_1,x_2,t\right)\right] \exp\left(Ik_p x_p\right)\ell^2 d\mathbf{k} = \\
&= \frac{1}{\left(2\pi\right)^2}\int_{-\infty}^{+\infty}\int_{-\infty}^{+\infty} \mathcal{Z}\left[f_{i_1,\ldots,i_{n/2}}\right]\Big|_{z_j=\exp\left(In_r^{(j)} k_r \ell\right)} \exp\left(Ik_p x_p\right)\ell^2 d\mathbf{k} = \\
&= \frac{1}{\left(2\pi\right)^2}\sum_{-\infty}^{+\infty}\ldots\sum_{-\infty}^{+\infty} f_{i_1,\ldots,i_{n/2}i}\left(t\right)\int_{-\infty}^{+\infty}\int_{-\infty}^{+\infty}\left[\prod_{j=1}^{n/2}\exp\left(-Ii_j n_r^{(j)} k_r \ell\right)\right]\exp\left(Ik_p x_p\right)\ell^2 d\mathbf{k}
\end{aligned} \tag{41}
$$

with $p = 1,2$ .

The inverse Fourier transform of equation (39) provides the following integral-differential equation

$$
\mathcal{F}^{-1}\left[\begin{pmatrix} \alpha_{i_1-1,\ldots,i_{n/2}} \exp\left(-I\ n_r^{(1)} k_r \ell\right) + \alpha_{i_1+1,\ldots,i_{n/2}} \exp\left(I\ n_r^{(1)} k_r \ell\right) + \alpha_{i_1,\ldots,i_{n/2}} + \ldots\ldots + \\ +\alpha_{i_1,\ldots,i_{n/2}-1} \exp\left(-I\ n_r^{(n/2)} k_r \ell\right) + \alpha_{i_1,\ldots,i_{n/2}+1} \exp\left(I\ n_r^{(n/2)} k_r \ell\right) \end{pmatrix} \hat{\Psi}\left(k_1,k_2,t\right)\right] + \\ + \mathcal{F}^{-1}\left[\hat{F}\left(k_1,k_2,t\right)\right] = I_\psi \mathcal{F}^{-1}\left[\ddot{\hat{\Psi}}\left(k_1,k_2,t\right)\right], \tag{42}
$$

that is specialized as

$$
\frac{1}{(2\pi)^2}\int_{-\infty}^{+\infty}\int_{-\infty}^{+\infty}\left\{\begin{bmatrix}\alpha_{i_1-1,\dots,i_{n/2}}\exp\left(-I\ n_r^{(1)}k_r\ell\right)+\alpha_{i_1+1,\dots,i_{n/2}}\exp\left(I\ n_r^{(1)}k_r\ell\right)+\alpha_{i_1,\dots,i_{n/2}}+\dots\dots+\\+\alpha_{i_1,\dots,i_{n/2}-1}\exp\left(-I\ n_r^{(n/2)}k_r\ell\right)+\alpha_{i_1,\dots,i_{n/2}+1}\exp\left(I\ n_r^{(n/2)}k_r\ell\right)\end{bmatrix}\times\atop \times\mathcal{F}\left[\Psi\left(x_1,x_2,t\right)\right]\exp\left(Ik_px_p\right)\ell^2 d\mathbf{k}\right\}+
$$

$$
+\frac{1}{(2\pi)^2}\int_{-\infty}^{+\infty}\int_{-\infty}^{+\infty}\mathcal{F}\left[F\left(x_1,x_2,t\right)\right]\exp\left(Ik_px_p\right)\ell^2 d\mathbf{k}= \tag{43}
$$

$$
=I_\psi\frac{1}{(2\pi)^2}\int_{-\infty}^{+\infty}\int_{-\infty}^{+\infty}\mathcal{F}\left[\ddot{\Psi}\left(x_1,x_2,t\right)\right]\exp\left(Ik_px_p\right)\ell^2 d\mathbf{k}\ .
$$

Recalling the notion of pseudo-differential operators introduced in Section 2, the integral-differential equation may be written as a pseudo-differential equation in the form

$$
\begin{bmatrix}\alpha_{i_1-1,\dots,i_{n/2}}\exp\left(-\ell\ n_r^{(1)}D_r\right)+\alpha_{i_1+1,\dots,i_{n/2}}\exp\left(\ell\ n_r^{(1)}D_r\right)+\alpha_{i_1,\dots,i_{n/2}}+\dots\dots+\\+\alpha_{i_1,\dots,i_{n/2}-1}\exp\left(-\ell\ n_r^{(n/2)}D_r\right)+\alpha_{i_1,\dots,i_{n/2}+1}\exp\left(\ell\ n_r^{(n/2)}D_r\right)\end{bmatrix}\Psi\left(x_1,x_2,t\right)+ \tag{44}
$$
$$
+F\left(x_1,x_2,t\right)=I_\psi\ddot{\Psi}\left(x_1,x_2,t\right),
$$

where the differential operator $D_r\left(\cdot\right)=\dfrac{\partial}{\partial x_r}\left(\cdot\right)$ is introduced.

Following the approach presented in Section 2 in order to obtain a thermodynamically consistent continuum equivalent to the lattice, a proper mapping is now introduced relating the transformed displacement $\hat{\Psi}\left(k_1,k_2,t\right)$ in the Fourier space to an auxiliary regularizing continuum field $\hat{\Psi}^R\left(k_1,k_2,t\right)$ in the same space as follows

$$
\hat{\Psi}^R\left(k_1,k_2,t\right)\doteq\prod_{j=1}^{n/2}\frac{\exp\left(In_r^{(j)}k_r\ell\right)-\exp\left(-In_r^{(j)}k_r\ell\right)}{2I\ n_r^{(j)}k_r\ell}\ \hat{\Psi}\left(k_1,k_2,t\right). \tag{45}
$$

By substituting mapping (45) in the equation of motion written in the transformed space (39) one obtains

$$
\begin{pmatrix}\alpha_{i_1-1,\dots,i_{n/2}}\exp\left(-I\ n_r^{(1)}k_r\ell\right)+\alpha_{i_1+1,\dots,i_{n/2}}\exp\left(I\ n_r^{(1)}k_r\ell\right)+\alpha_{i_1,\dots,i_{n/2}}+\dots\dots+\\+\alpha_{i_1,\dots,i_{n/2}-1}\exp\left(-I\ n_r^{(n/2)}k_r\ell\right)+\alpha_{i_1,\dots,i_{n/2}+1}\exp\left(I\ n_r^{(n/2)}k_r\ell\right)\end{pmatrix}\times
$$
$$
\times\prod_{j=1}^{n/2}\frac{2I\ n_r^{(j)}k_r\ell}{\exp\left(In_r^{(j)}k_r\ell\right)-\exp\left(-In_r^{(j)}k_r\ell\right)}\hat{\Psi}^R\left(k_1,k_2,t\right)+\hat{F}\left(k_1,k_2,t\right) \tag{46}
$$
$$
=I_\Psi\prod_{j=1}^{n/2}\frac{2I\ n_r^{(j)}k_r\ell}{\exp\left(In_r^{(j)}k_r\ell\right)-\exp\left(-In_r^{(j)}k_r\ell\right)}\ddot{\hat{\Psi}}^R\left(k_1,k_2,t\right).
$$

By inverse transform of equation (46) and remembering the notation $\hat{\Psi}^R(k_1,k_2,t)=\mathcal{F}\left[\Psi^R(x_1,x_2,t)\right]$ the following integral-differential equation is obtained

$$\mathcal{F}^{-1}\left[\begin{pmatrix}\alpha_{i_1-1,\dots,i_{n/2}}\exp\left(-I\ n_r^{(1)}k_r\ell\right)+\alpha_{i_1+1,\dots,i_{n/2}}\exp\left(I\ n_r^{(1)}k_r\ell\right)+\alpha_{i_1,\dots,i_{n/2}}+\dots\dots+\\+\alpha_{i_1,\dots,i_{n/2}-1}\exp\left(-I\ n_r^{(n/2)}k_r\ell\right)+\alpha_{i_1,\dots,i_{n/2}+1}\exp\left(I\ n_r^{(n/2)}k_r\ell\right)\end{pmatrix}\times \atop \times\prod_{j=1}^{n/2}\frac{2I\ n_r^{(j)}k_r\ell}{\exp\left(In_r^{(j)}k_r\ell\right)-\exp\left(-In_r^{(j)}k_r\ell\right)}\mathcal{F}\left[\Psi^R(x_1,x_2,t)\right]\right]+ \tag{47}$$

$$+\mathcal{F}^{-1}\left[\hat{F}(k_1,k_2,t)\right]=I_\Psi\mathcal{F}^{-1}\left[\prod_{j=1}^{n/2}\frac{2I\ n_r^{(j)}k_r\ell}{\exp\left(In_r^{(j)}k_r\ell\right)-\exp\left(-In_r^{(j)}k_r\ell\right)}\mathcal{F}\left[\ddot{\Psi}^R(x_1,x_2,t)\right]\right].$$

By the concept of pseudo-differential operator, the integral-differential field equation of motion may be written in an equivalent form

$$\begin{pmatrix}\alpha_{i_1-1,\dots,i_{n/2}}\exp\left(-\ell\ n_r^{(1)}D_r\right)+\alpha_{i_1+1,\dots,i_{n/2}}\exp\left(\ell\ n_r^{(1)}D_r\right)+\alpha_{i_1,\dots,i_{n/2}}+\dots\dots+\\+\alpha_{i_1,\dots,i_{n/2}-1}\exp\left(-\ell\ n_r^{(n/2)}D_r\right)+\alpha_{i_1,\dots,i_{n/2}+1}\exp\left(\ell\ n_r^{(n/2)}D_r\right)\end{pmatrix}\times$$
$$\times\prod_{j=1}^{n/2}\frac{2I\ n_r^{(j)}k_r\ell}{\exp\left(In_r^{(j)}k_r\ell\right)-\exp\left(-In_r^{(j)}k_r\ell\right)}\Psi^R(x_1,x_2,t)+F(x_1,x_2,t)= \tag{48}$$
$$=I_\Psi\prod_{j=1}^{n/2}\frac{2I\ n_r^{(j)}k_r\ell}{\exp\left(In_r^{(j)}k_r\ell\right)-\exp\left(-In_r^{(j)}k_r\ell\right)}\ddot{\Psi}^R(x_1,x_2,t)\quad .$$

Here two new pseudo-differential operators are introduced, namely

$$P_3(D_1,D_2)=$$
$$=\begin{pmatrix}\alpha_{i_1-1,\dots,i_{n/2}}\exp\left(-\ell\ n_r^{(1)}D_r\right)+\alpha_{i_1+1,\dots,i_{n/2}}\exp\left(\ell\ n_r^{(1)}D_r\right)+\alpha_{i_1,\dots,i_{n/2}}+\dots\dots+\\+\alpha_{i_1,\dots,i_{n/2}-1}\exp\left(-\ell\ n_r^{(n/2)}D_r\right)+\alpha_{i_1,\dots,i_{n/2}+1}\exp\left(\ell\ n_r^{(n/2)}D_r\right)\end{pmatrix}\times \tag{49}$$
$$\times\prod_{j=1}^{n/2}\frac{2\ \ell n_r^{(j)}D_r}{\exp\left(\ell\ n_r^{(j)}D_r\right)-\exp\left(-\ell\ n_r^{(j)}D_r\right)},$$

and

$$P_4(D_1,D_2)=\prod_{j=1}^{n/2}\frac{2\ \ell n_r^{(j)}D_r}{\exp\left(\ell\ n_r^{(j)}D_r\right)-\exp\left(-\ell\ n_r^{(j)}D_r\right)}. \tag{50}$$

It is worth to note that from an inverse Fourier transform of the mapping (45) and remembering the definition (38) the remarkable relation is obtained between the directional derivative of the regularized continuum field and the effective nodal displacement in the discrete given model

$$\left.\left(\prod_{j=1}^{n/2}\left(n_r^{(j)}D_r\right)\ddot{\Psi}^R\left(x_1,x_2,t\right)\right)\right|_{\mathbf{x}_0}=\prod_{j=1}^{n/2}\frac{\exp\left(\ell\, n_r^{(j)}D_r\right)-\exp\left(-\ell\, n_r^{(j)}D_r\right)}{2\ell}\psi_{i_1,\dots,i_{n/2}}, \tag{51}$$

where $\exp\left(\ell\, n_r^{(j)}D_r\right)$ plays the role of shift operator according to Jordan (1965), Rota *et al.* (1973), Kelley and Peterson (2001), Bacigalupo and Gambarotta (2019) and the position vector of the reference node is $\mathbf{x}_0=n_r^{(1)}\ell\, m_1+....+n_r^{(n/2)}\ell\, m_{n/2}$ with $m_1,....\, m_{n/2}\in\mathbb{Z}$.

In analogy to the approach presented in Section 2, higher order differential field equations may be obtained through a multivariate Taylor expansions in the $\mathbf{k}$ variable of the two kernels of the integral-differential equation (47) or, alternatively, through a formal Taylor expansion in $D$ variable of the pseudo-differential operators in equation (48).

## 4. Illustrative examples

In order to show the reliability and the validity limits of the here proposed enhanced continualization approach applied to lattice-like materials, let us consider two nets, made up of massless strings of equal length $\ell$ with nodal mass $m$ and subjected to a pre-stress N. The first one is a square net (Figure 4.a) while the second one is a triangular net (figure 4.b).

### *4.1 Square cable net*

The equation of motion of the reference node of the Lagrangian model reads

$$\psi_{i_1-1,i_2}+\psi_{i_1+1,i_2}+\psi_{i_1,i_2-1}+\psi_{i_1,i_2+1}-4\psi_{i_1,i_2}+f_{i_1,i_2}=I_\psi\ddot{\psi}_{i_1,i_2} \tag{52}$$

and the dispersion equation representing the dimensionless angular frequency in terms of the wave vector components is

$$\omega\sqrt{I_\psi}=\sqrt{4-2\cos\left(k_1\ell\right)-2\cos\left(k_2\ell\right)}\quad . \tag{53}$$

By applying the standard continualization to equation (52), i.e. by representing equation (44) specialized to the case of square cable-net as a formal Taylor series in the $D$ variable and after a second order truncation (*2nd order standard continualization*) the classical equation of the transverse motion of the isotropic membrane is found

$$\ell^2\Delta\Psi+f=I_\psi\ddot{\Psi}\quad . \tag{54}$$

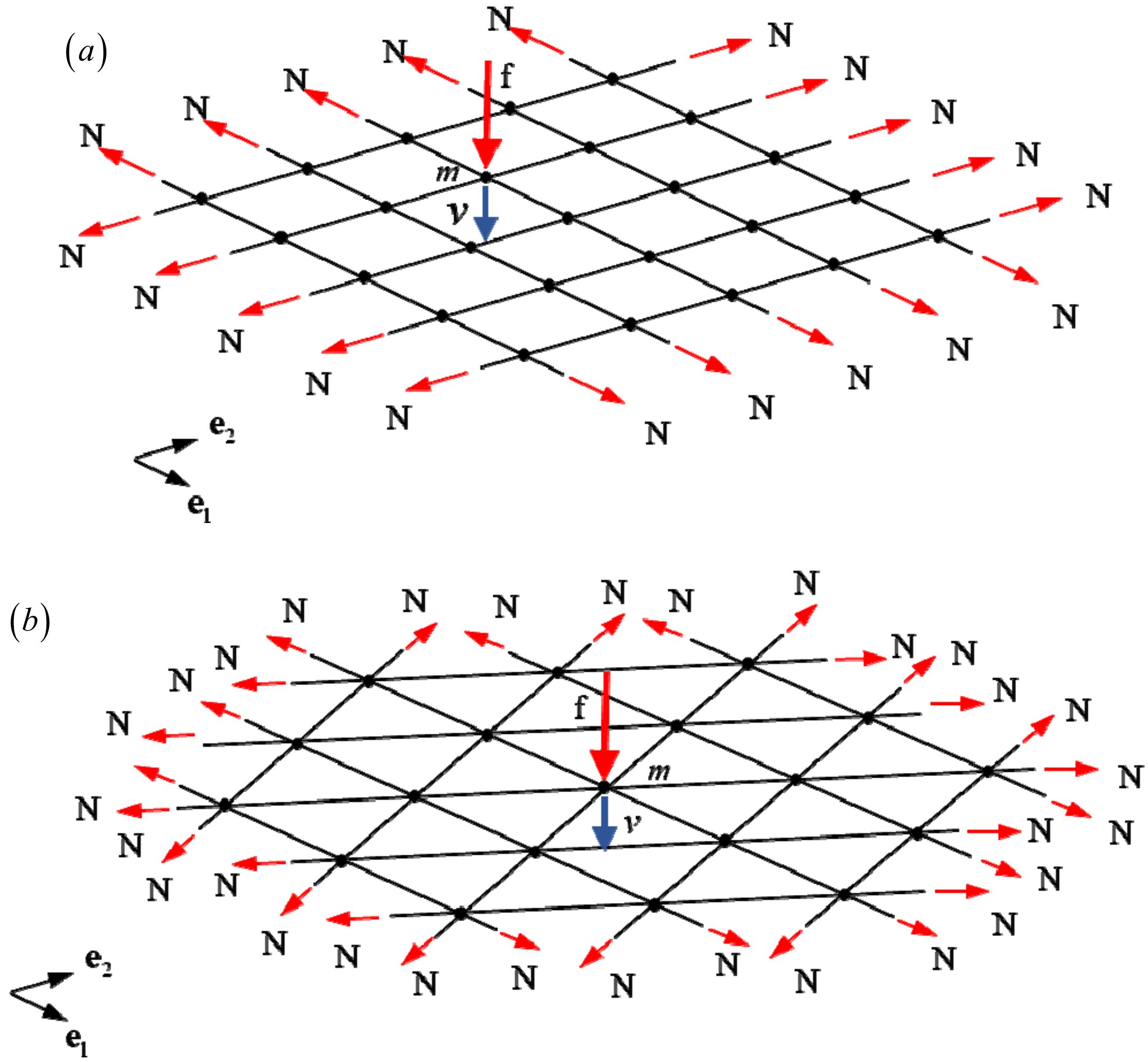

Figure 4. Square cable net (a); triangular cable net (b).

Analogously, if a fourth order truncation (*4^th^ order standard continualization*) is applied, the following higher order governing equation is obtained

$$\ell^2 \Delta\Psi + \frac{\ell^4}{12}\left(\frac{\partial^4\Psi}{\partial x_1^4} + \frac{\partial^4\Psi}{\partial x_2^4}\right) + f = I_\psi \ddot{\Psi} \qquad (55)$$

whose Lagrangian functional turns out to be characterized by a non-positive defined potential energy density. Moreover, the structure of equation (55) replicates the two-dimensional cubic symmetry of the square cable-net. Following the enhanced continualization here proposed, the pseudo-differential operators (49) and (50) have to be expanded in the following formal Taylor series

$$P_3\left(D_1,D_2\right)=\left(D_1{}^2+D_2{}^2\right)\ell^2-\frac{1}{12}\left(D_1{}^4+4D_1{}^2D_2{}^2+D_2{}^4\right)\ell^4+ \\ +\frac{1}{120}\left(D_1{}^6+4D_1{}^2D_2{}^4+4D_1{}^4D_2{}^2+D_2{}^6\right)\ell^6+O\left(\ell^8\right) \quad , \tag{56}$$

$$P_4\left(D_1,D_2\right)=1-\frac{1}{6}\left(D_1^2+D_2^2\right)\ell^2+\frac{7}{360}\left(D_1^4+10D_1^2D_2^2+D_2^4\right)\ell^4+ \\ -\frac{31}{15120}\left(D_1^6+49D_1^2D_2^4+49D_1^4D_2^2+D_2^6\right)\ell^6+O\left(\ell^8\right) \quad . \tag{57}$$

This allows the integral-differential field equation (48) to be approximated as a partial differential equation at a required order. If a second order truncation is applied (*2nd order enhanced continualization*) the following isotropic non-local governing equation is found

$$\ell^2\Delta\Psi^R=I_\psi\left(\ddot{\Psi}^R-\frac{1}{6}\ell^2\Delta\ddot{\Psi}^R\right) \quad , \tag{58}$$

that is formally similar equation to the one derived by Andrianov and Awrejcewicz (2008) whose kinetic and elastic potential energy density are positive defined. If a fourth order truncation (*4th enhanced continualization*) is carried out the following non-local higher order equation is found

$$\ell^2\Delta\Psi^R-\frac{\ell^4}{12}\left(\frac{\partial^4\Psi^R}{\partial x_1^4}+4\frac{\partial^4\Psi^R}{\partial x_1^2\partial x_2^2}+\frac{\partial^4\Psi^R}{\partial x_2^4}\right)= \\ =I_\psi\left[\ddot{\Psi}^R-\frac{1}{6}\ell^2\Delta\ddot{\Psi}^R+\frac{7}{360}\ell^4\left(\frac{\partial^4\ddot{\Psi}^R}{\partial x_1^4}+10\frac{\partial^4\ddot{\Psi}^R}{\partial x_1^2\partial x_2^2}+\frac{\partial^4\ddot{\Psi}^R}{\partial x_2^4}\right)\right] \quad . \tag{59}$$

which replicates the two-dimensional cubic symmetry of the square cable-net. The potential energy density of the equivalent continuum is a positive defined and convex function

$$e\left(\frac{\partial\Psi^R}{\partial x_\alpha},\frac{\partial^2\Psi^R}{\partial x_\alpha\partial x_\beta}\right)=\frac{1}{2}\ell^2\left[\left(\frac{\partial\Psi^R}{\partial x_1}\right)^2+\left(\frac{\partial\Psi^R}{\partial x_2}\right)^2\right]+ \\ +\frac{1}{24}\ell^4\left[\left(\frac{\partial^2\Psi^R}{\partial x_1{}^2}\right)^2+2\left(\frac{\partial^2\Psi^R}{\partial x_1\partial x_2}\right)^2+2\left(\frac{\partial^2\Psi^R}{\partial x_2\partial x_1}\right)^2+\left(\frac{\partial^2\Psi^R}{\partial x_2{}^2}\right)^2\right] \quad . \tag{60}$$

Accordingly, the first order stress tensor components $\tau_\alpha=\frac{\partial e}{\partial\Psi_{,\alpha}}$, the second order stress tensor components $\tau_{\alpha\beta}=\frac{\partial e}{\partial\Psi_{,\alpha\beta}}$ and the total stress components $\sigma_\alpha=\tau_\alpha-\frac{\partial\tau_{\alpha\beta}}{\partial x_\beta}$ may be defined, respectively (the comma here represents the spatial partial derivative). The constitutive equations of the homogeneous continuum are

$$\tau_1 = \ell^2 \frac{\partial \Psi^R}{\partial x_1}, \quad \tau_2 = \ell^2 \frac{\partial \Psi^R}{\partial x_2} \quad , \tag{61}$$

$$\tau_{11} = \frac{\ell^4}{12}\frac{\partial^2 \Psi^R}{\partial x_1{}^2}, \quad \tau_{22} = \frac{\ell^4}{12}\frac{\partial^2 \Psi^R}{\partial x_2{}^2}, \quad \tau_{12} = \tau_{21} = \frac{\ell^4}{6}\frac{\partial^2 \Psi^R}{\partial x_1 \partial x_2} \quad , \tag{62}$$

$$\begin{aligned} \sigma_1 &= \tau_1 - \frac{\partial \tau_{11}}{\partial x_1} - \frac{\partial \tau_{12}}{\partial x_2} = \ell^2 \frac{\partial \Psi^R}{\partial x_1} - \frac{\ell^4}{12}\frac{\partial^3 \Psi^R}{\partial x_1{}^3} - \frac{\ell^4}{6}\frac{\partial^2 \Psi^R}{\partial x_1 \partial x_2{}^2} \\ \sigma_2 &= \tau_2 - \frac{\partial \tau_{21}}{\partial x_1} - \frac{\partial \tau_{22}}{\partial x_2} = \ell^2 \frac{\partial \Psi^R}{\partial x_2} - \frac{\ell^4}{6}\frac{\partial^2 \Psi^R}{\partial x_1{}^2 \partial x_2} - \frac{\ell^4}{12}\frac{\partial^3 \Psi^R}{\partial x_2{}^3} \quad . \end{aligned} \tag{63}$$

Moreover, also the kinetic energy density is a positive defined and convex function

$$\begin{aligned} &\kappa\left(\dot{\Psi}^R, \frac{\partial \dot{\Psi}^R}{\partial x_\alpha}, \frac{\partial^2 \dot{\Psi}^R}{\partial x_\alpha \partial x_\beta}\right) = \\ &= \frac{1}{2} I_\psi \left\{ \begin{array}{l} \left(\dot{\Psi}^R\right)^2 + \frac{1}{6}\ell^2 \left[\left(\frac{\partial\ \dot{\Psi}^R}{\partial x_1}\right)^2 + \left(\frac{\partial\ \dot{\Psi}^R}{\partial x_2}\right)^2\right] + \\ + \frac{7}{360}\ell^4 \left[\left(\frac{\partial^2 \dot{\Psi}^R}{\partial x_1{}^2}\right)^2 + 5\left(\frac{\partial^2 \dot{\Psi}^R}{\partial x_1 \partial x_2}\right)^2 + 5\left(\frac{\partial^2 \dot{\Psi}^R}{\partial x_2 \partial x_1}\right)^2 + \left(\frac{\partial^2 \dot{\Psi}^R}{\partial x_2{}^2}\right)^2\right] \end{array} \right\} . \end{aligned} \tag{64}$$

A local $\upsilon = \frac{\partial \kappa}{\partial \dot{\Psi}^R}$ and non-local inertia terms $\upsilon_\alpha = \frac{\partial \kappa}{\partial \dot{\Psi}^R_{,\alpha}}$ and $\upsilon_{\alpha\beta} = \frac{\partial \kappa}{\partial \dot{\Psi}^R_{,\alpha\beta}}$ may be identified and the following inertial constitutive equations are obtained

$$\upsilon = I_\psi \dot{\Psi}^R, \quad \upsilon_1 = \frac{1}{6} I_\psi \ell^2 \frac{\partial \dot{\Psi}^R}{\partial x_1}, \quad \upsilon_2 = \frac{1}{6} I_\psi \ell^2 \frac{\partial \dot{\Psi}^R}{\partial x_2} \tag{65}$$

$$\upsilon_{11} = \frac{7}{360}\ell^4 \frac{\partial^2 \dot{\Psi}^R}{\partial x_1{}^2}, \quad \upsilon_{22} = \frac{7}{360}\ell^4 \frac{\partial^2 \dot{\Psi}^R}{\partial x_2{}^2}, \quad \upsilon_{12} = \upsilon_{21} = \frac{35}{360}\ell^4 \frac{\partial^2 \dot{\Psi}^R}{\partial x_1 \partial x_2} \quad . \tag{66}$$

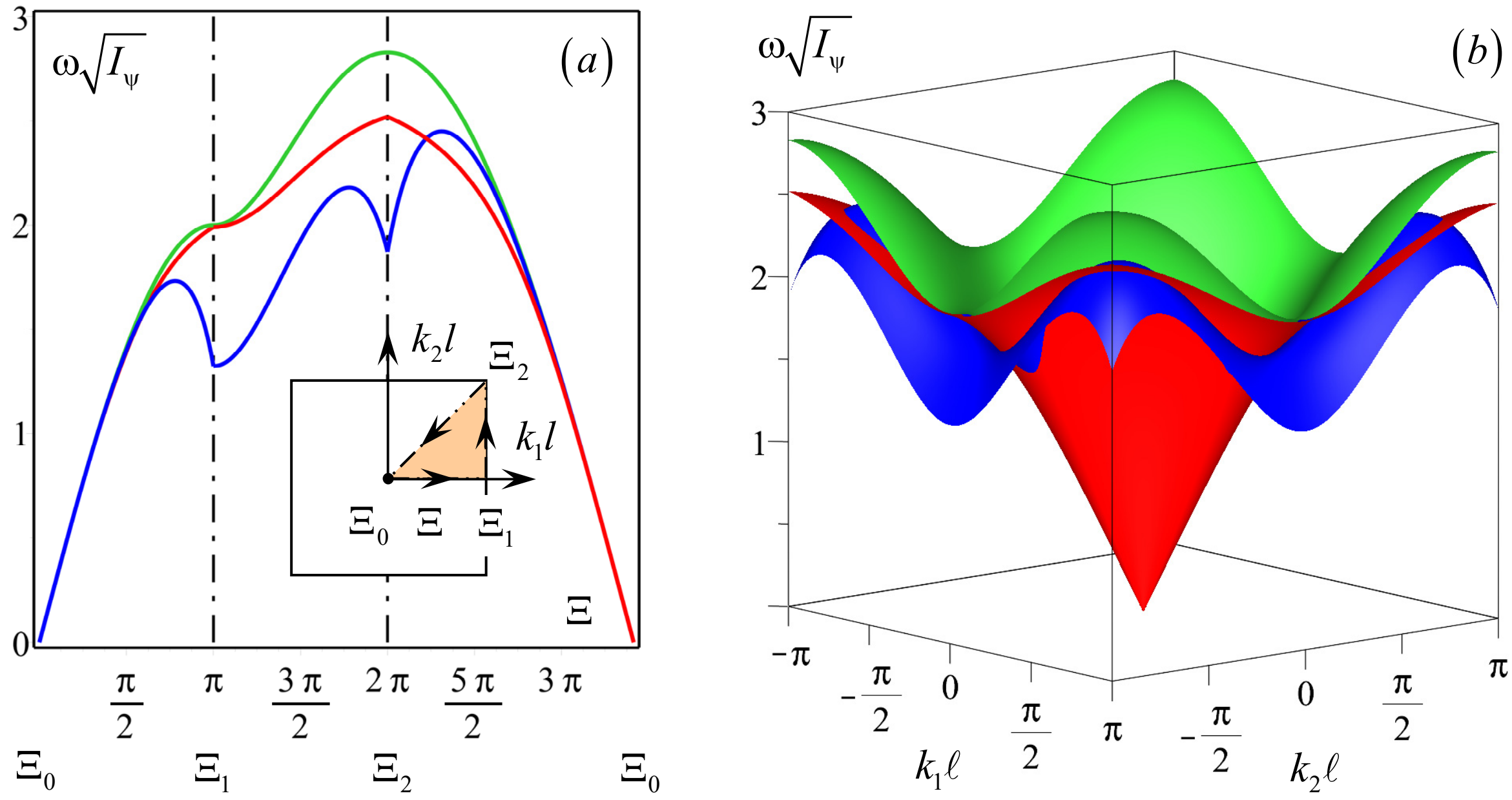

Figure 5. Lagrangian model (green line), 4th order standard continualization (blue line), 4th order enhanced continualization (red line) of the square pre-stressed cable net.

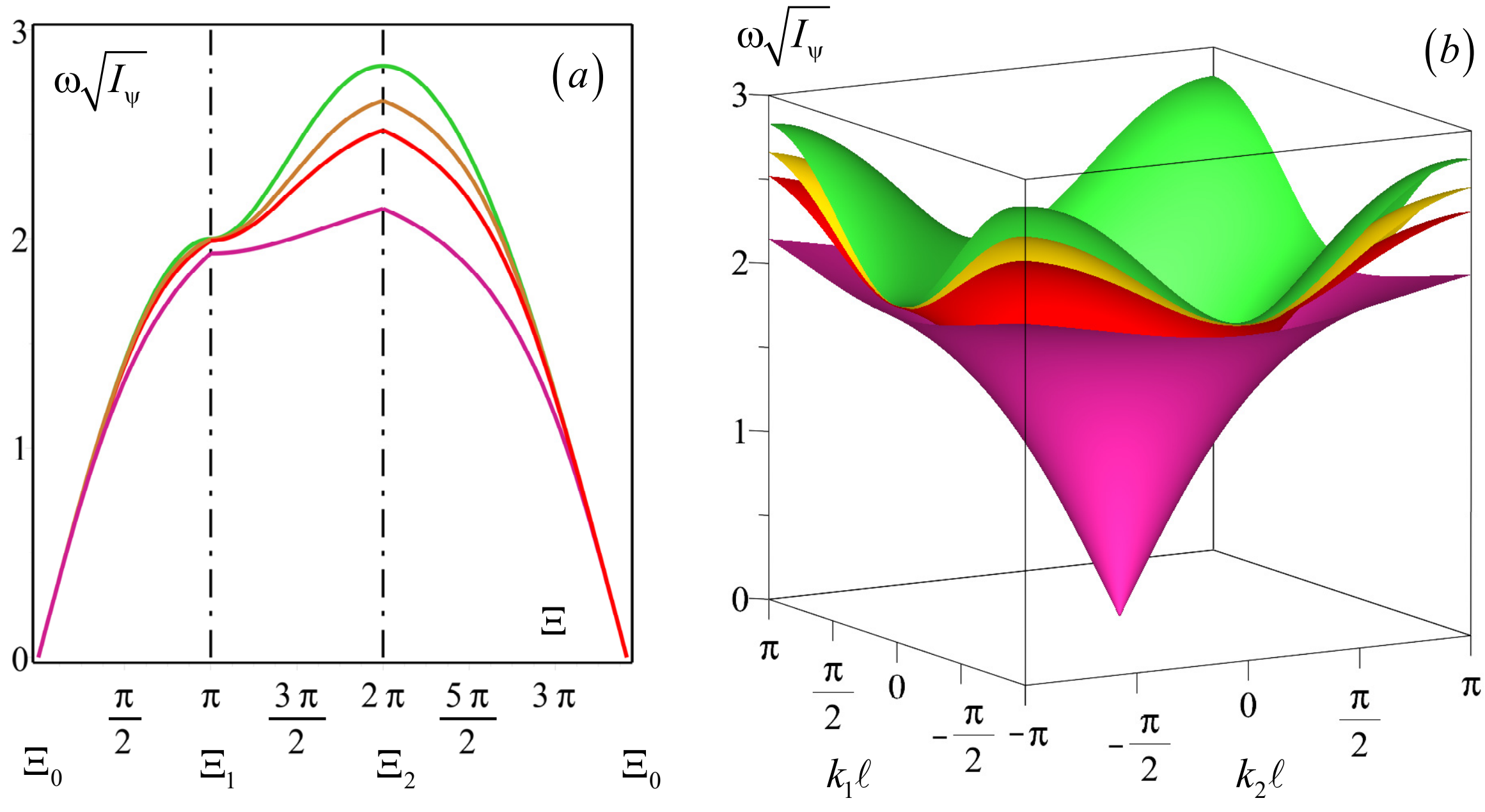

Figure 6. Lagrangian model (green line), 2nd order enhanced continualization (violet line), 4th order enhanced continualization (red line), 6th order enhanced continualization (yellow line) of the square pre-stressed cable net.

The dispersive function of the square pre-stressed cable net obtained by the discrete model is shown in Figure 5.a and compared with the corresponding functions obtained via standard and enhanced continualization, respectively. The dimensionless angular frequency $\omega\sqrt{I_\psi}$ is expressed as function of the arch length $\Xi$ measured on the closed polygonal curve $\Upsilon$ with vertices identified by the values $\Xi_j,\ \ j=0,1,2$, encompassing the first irreducible Brillouin zone. It is worth to note that dispersive function obtained via 4th order enhanced continualization (red line – fourth order truncation in $\ell$ ) turns out to be in good agreement with the actual corresponding one (green line). This agreement is found for the 4th order standard continualization (blue line) in the long wavelength regime, while for shorter wavelength a lesser accuracy is found, with prevailing negative group velocity. Figure 5.b provides the dispersive surfaces for the same models in the whole first Brillouin zone. Regions in which a negative group velocity associated to the 4th order standard continualization are shown. In Figure 6 the convergence of the dispersive functions obtained via enhanced continualization to the actual one is shown when increasing the continualization order.

*4.2 Triangular cable net*

The equation of motion of the reference node reads

$$\psi_{i_1-1,i_2,i_3}+\psi_{i_1+1,i_2,i_3}+\psi_{i_1,i_2-1,i_3}+\psi_{i_1,i_2+1,i_3}+\psi_{i_1,i_2,i_3-1}+\psi_{i_1,i_2,i_3+1}-6\psi_{i_1,i_2,i_3}+f_{i_1,i_2,i_3}=I_\psi\ddot{\psi}_{i_1,i_2,i_3} \quad (67)$$

and the dispersion function takes the following form

$$\omega\sqrt{I_\psi}=\sqrt{6-2\cos\left(k_1\ell\right)-4\cos\left(\frac{1}{2}k_1\ell\right)\cos\left(\frac{\sqrt{3}}{2}k_2\ell\right)}\quad . \quad (68)$$

The equation of motion of the equivalent continuum obtained via 2nd order standard continualization corresponds to the governing equation of the isotropic membrane

$$\frac{3}{2}\ell^2\Delta\Psi+f=I_\psi\ddot{\Psi}\quad , \quad (69)$$

while the 4th order standard continualization provides the non-local equation of motion of the equivalent continuum

$$\frac{3}{2}\ell^2\Delta\Psi+\frac{3}{32}\ell^4\left(\frac{\partial^4\Psi}{\partial x_1^4}+\frac{\partial^4\Psi}{\partial x_1^2\partial x_2^2}+\frac{\partial^4\Psi}{\partial x_2^4}\right)+f=I_\psi\ddot{\Psi} \quad (70)$$

having a non-positive defined potential energy density. The structure of equation (70) presents two-dimensional cubic symmetry. When applying the enhanced continualization here proposed, the formal expansion of the pseudo-differential operators (49) and (50) may be expressed as power series

$$P_1(D_1,D_2)=\frac{3}{2}\left(D_1{}^2+D_2{}^2\right)\ell^2-\frac{9}{32}\left(D_1{}^4+2D_1{}^2D_2{}^2+D_2{}^4\right)\ell^4+$$
$$+\frac{137}{3840}\left(D_1{}^6+\frac{393}{137}D_1{}^4D_2{}^2+\frac{423}{137}D_1{}^2D_2{}^4+\frac{135}{137}D_2{}^6\right)\ell^6+O\left(\ell^8\right)\quad, \tag{71}$$

$$P_2(D_1,D_2)=1-\frac{1}{4}\left(D_1^2+D_2^2\right)\ell^2+\frac{3}{80}\left(D_1^4+2D_1^2D_2^2+D_2^4\right)\ell^4+$$
$$-\frac{137}{30240}\left(D_1^6+\frac{393}{137}D_1^4D_2^2+\frac{423}{137}D_1^2D_2^4+\frac{135}{137}D_2^6\right)\ell^6+O\left(\ell^8\right)\quad. \tag{72}$$

In case of 2nd order enhanced continualization the isotropic non-local partial differential equation is obtained

$$\frac{3}{2}\ell^2\Delta\Psi^R=I_\psi\left(\ddot{\Psi}^R-\frac{1}{4}\ell^2\Delta\ddot{\Psi}^R\right)=0\quad, \tag{73}$$

while performing a 4th enhanced continualization the resulting isotropic governing equation reads

$$\frac{3}{2}\ell^2\Delta\Psi^R-\frac{9}{32}\left(\frac{\partial^4\Psi^R}{\partial x_1^4}+2\frac{\partial^4\Psi^R}{\partial x_1^2\partial x_2^2}+\frac{\partial^4\Psi^R}{\partial x_2^4}\right)\ell^4=$$
$$=I_\psi\left[\ddot{\Psi}^R-\frac{1}{4}\ell^2\Delta\ddot{\Psi}^R+\frac{3}{80}\ell^4\left(\frac{\partial^4\ddot{\Psi}^R}{\partial x_1^4}+2\frac{\partial^4\ddot{\Psi}^R}{\partial x_1^2\partial x_2^2}+\frac{\partial^4\ddot{\Psi}^R}{\partial x_2^4}\right)\right]\quad. \tag{74}$$

The potential energy density of the equivalent continuum is a positive defined and convex function written as

$$e\left(\frac{\partial\Psi^R}{\partial x_\alpha},\frac{\partial^2\Psi^R}{\partial x_\alpha\partial x_\beta}\right)=\frac{3}{4}\ell^2\left[\left(\frac{\partial\Psi^R}{\partial x_1}\right)^2+\left(\frac{\partial\Psi^R}{\partial x_2}\right)^2\right]+$$
$$+\frac{9}{64}\ell^4\left[\left(\frac{\partial^2\Psi^R}{\partial x_1{}^2}\right)^2+\left(\frac{\partial^2\Psi^R}{\partial x_1\partial x_2}\right)^2+\left(\frac{\partial^2\Psi^R}{\partial x_2\partial x_1}\right)^2+\left(\frac{\partial^2\Psi^R}{\partial x_2{}^2}\right)^2\right]\quad. \tag{75}$$

The first, second order stress components and the total stress components are, respectively, obtained in the form

$$\tau_1=\frac{3}{2}\ell^2\frac{\partial\Psi^R}{\partial x_1},\quad \tau_2=\frac{3}{2}\ell^2\frac{\partial\Psi^R}{\partial x_2} \tag{76}$$

$$\tau_{11}=\frac{9}{32}\ell^4\frac{\partial^2\Psi^R}{\partial x_1{}^2},\qquad \tau_{22}=\frac{9}{32}\ell^4\frac{\partial^2\Psi^R}{\partial x_2{}^2},\qquad \tau_{12}=\tau_{21}=\frac{9}{32}\ell^4\frac{\partial^2\Psi^R}{\partial x_1\partial x_2} \tag{77}$$

$$
\begin{aligned}
\sigma_1 &= \frac{3}{2}\ell^2 \frac{\partial \Psi^R}{\partial x_1} - \frac{9}{32}\ell^4 \frac{\partial^3 \Psi^R}{\partial x_1^{\,3}} - \frac{9}{32}\ell^4 \frac{\partial^2 \Psi^R}{\partial x_1 \partial x_2^{\,2}} \\
\sigma_2 &= \frac{3}{2}\ell^2 \frac{\partial \Psi^R}{\partial x_2} - \frac{9}{32}\ell^4 \frac{\partial^2 \Psi^R}{\partial x_1^{\,2} \partial x_2} - \frac{9}{32}\ell^4 \frac{\partial^3 \Psi^R}{\partial x_2^{\,3}}
\end{aligned} \quad . \tag{78}
$$

Moreover, the kinetic energy density is a positive defined and convex function as well

$$
\begin{aligned}
&\kappa\left(\dot{\Psi}^R, \frac{\partial \dot{\Psi}^R}{\partial x_\alpha}, \frac{\partial^2 \dot{\Psi}^R}{\partial x_\alpha \partial x_\beta}\right) = \\
&= \frac{1}{2} I_\psi \left\{
\begin{aligned}
&\left(\dot{\Psi}^R\right)^2 + \frac{1}{4}\ell^2 \left[\left(\frac{\partial \dot{\Psi}^R}{\partial x_1}\right)^2 + \left(\frac{\partial \dot{\Psi}^R}{\partial x_2}\right)^2\right] + \\
&+ \frac{3}{80}\ell^4 \left[\left(\frac{\partial^2 \dot{\Psi}^R}{\partial x_1^{\,2}}\right)^2 + 5\left(\frac{\partial^2 \dot{\Psi}^R}{\partial x_1 \partial x_2}\right)^2 + 5\left(\frac{\partial^2 \dot{\Psi}^R}{\partial x_2 \partial x_1}\right)^2 + \left(\frac{\partial^2 \dot{\Psi}^R}{\partial x_2^{\,2}}\right)^2\right]
\end{aligned}
\right\} .
\end{aligned} \tag{79}
$$

The local inertial term $\upsilon = \frac{\partial t}{\partial \dot{\Psi}}$ and non-local inertia terms $\upsilon_\alpha = \frac{\partial t}{\partial \dot{\Psi}_{,\alpha}}$, $\upsilon_{\alpha\beta} = \frac{\partial t}{\partial \dot{\Psi}_{,\alpha\beta}}$ may be identified and the following inertial constitutive equations are obtained

$$
\upsilon = I_\psi \dot{\Psi}^R, \quad \upsilon_1 = \frac{1}{4} I_\psi \ell^2 \frac{\partial \dot{\Psi}^R}{\partial x_1}, \quad \upsilon_2 = \frac{1}{4} I_\psi \ell^2 \frac{\partial \dot{\Psi}^R}{\partial x_2} \tag{80}
$$

and

$$
\upsilon_{11} = \frac{1}{4}\ell^4 \frac{\partial^2 \dot{\Psi}^R}{\partial x_1^{\,2}}, \quad \upsilon_{22} = \frac{1}{4}\ell^4 \frac{\partial^2 \dot{\Psi}^R}{\partial x_2^{\,2}}, \quad \upsilon_{12} = \upsilon_{21} = \frac{3}{80}\ell^4 \frac{\partial^2 \dot{\Psi}^R}{\partial x_1 \partial x_2} \quad . \tag{81}
$$

The dispersive function of the triangular pre-stressed cable net from the lattice model is compared with those ones corresponding to the continualized models in Figure 7.a. The dimensionless angular frequency $\omega\sqrt{I_\psi}$ is expressed as function of the arch length $\Xi$ measured on the closed polygonal curve $\Upsilon$ with vertices identified by the values $\Xi_j$, $j = 0,1,2$, encompassing the first irreducible Brillouin zone. Also in this case the dispersive function obtained via 4[th] order enhanced continualization (red line) turns out to be in good agreement with the actual corresponding one (green line). As in the previous example, this agreement is found for the 4th order standard continualization (blue line) in the long wavelength regime, while for shorter wavelength a lesser accuracy is found, with prevailing negative group velocity together with short-wave instability and unlimited group velocity. Therefore, when considering the 4th order standard continualization, the homogenized model does not satisfy the Legendre-Hadamard condition in the entire first Brillouin zone with inconsistencies in the dynamic behaviour.

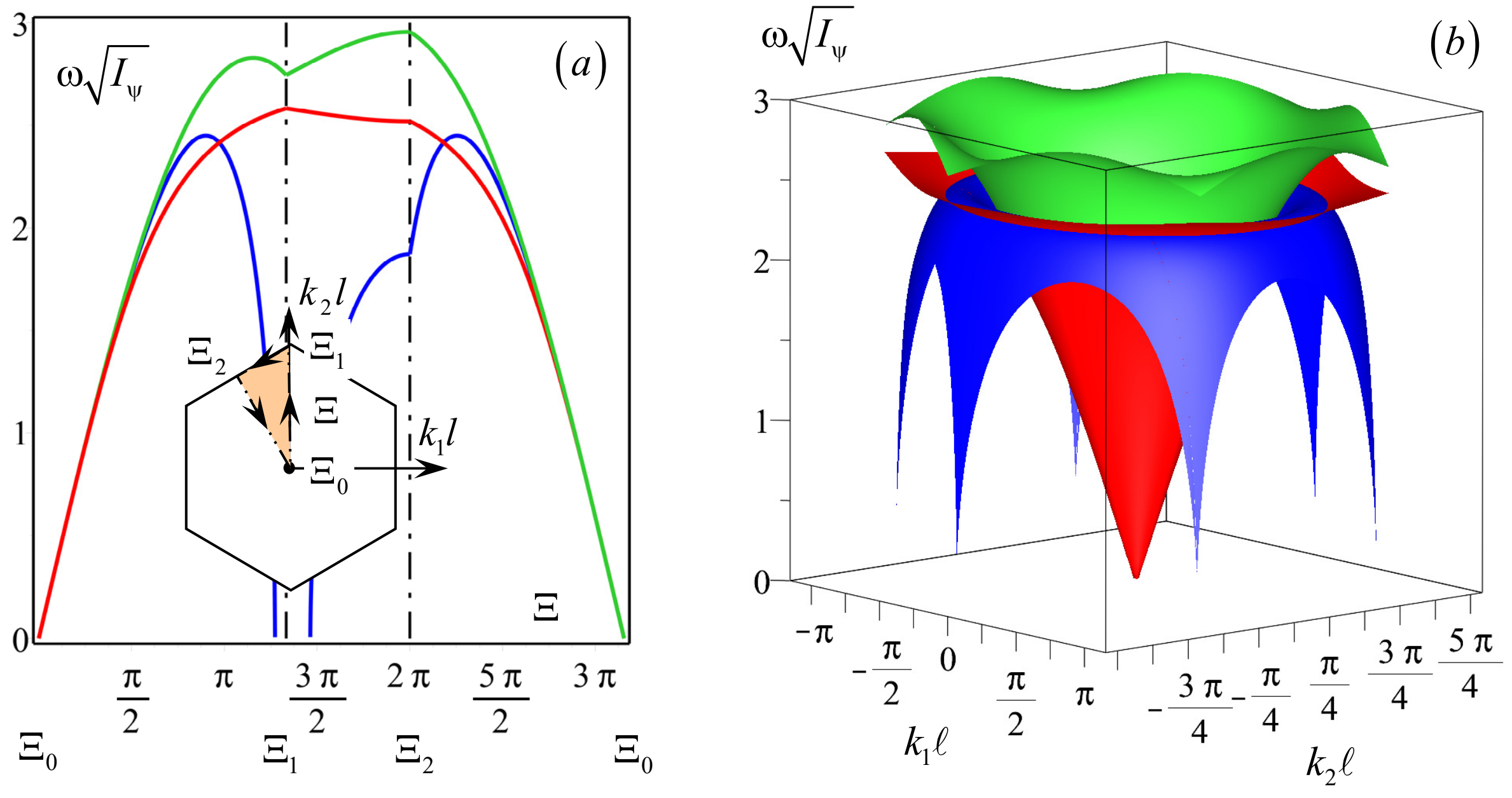

Figure 7. Lagrangian model (green line), 4th order standard continualization (blue line), 4th order enhanced continualization (red line) of the triangular pre-stressed cable net.

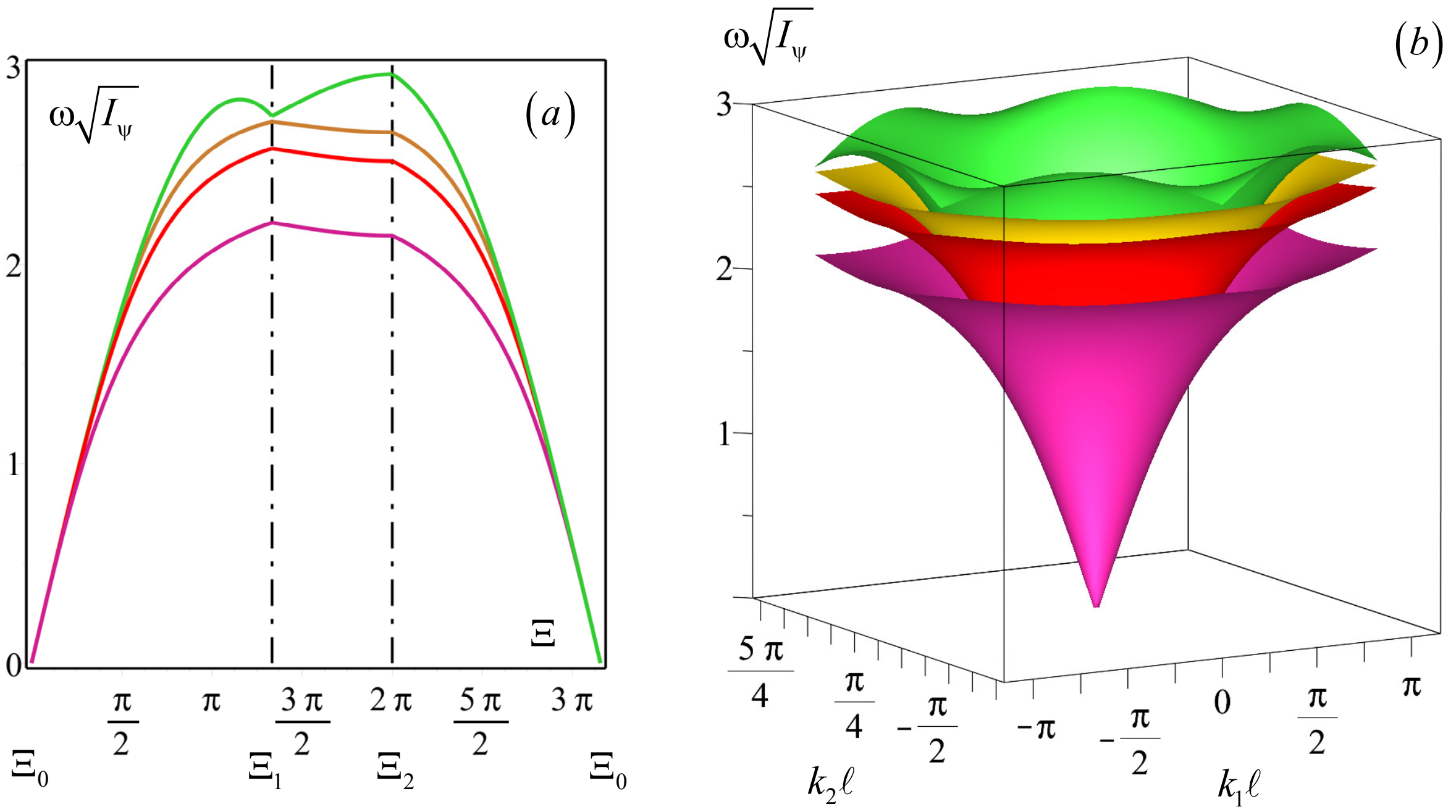

Figure 8. Lagrangian model (green line), 2nd order enhanced continualization (violet line), 4th order enhanced continualization (red line), 6th order enhanced continualization (gold line) of the triangular pre-stressed cable net.

Figure 7.b provides the dispersive surfaces for the same models in the whole first Brillouin zone. Regions in which a negative group velocity associated to the $4^{th}$ order standard continualization are shown. In Figure 8 the convergence of the dispersive functions obtained via enhanced continualization to the actual one is shown for increasing the continualization order. It is worth to note that the dispersive surfaces obtained via $6^{th}$ order enhanced continualization and higher order replicate the two-dimensional hexagonal symmetry of the triangular cable-net.

## 5. Conclusions

The standard continualization of lattice-like materials can lead, in general, to equivalent continuum models characterized by non-positive defined potential energy density. Therefore, inconsistencies in the dynamic behavior of these continuum models may result, such as short-wave instability, unlimited group velocity and non-fulfillment of the Legendre-Hadamard condition. To overcome these drawbacks, an energetically consistent technique has been proposed to identify non-local continua equivalent to one- and two-dimensional lattice systems having lamped mass at the nodes.

The equation of motion of the Lagrangian lattice model are transformed according to a unitary approach aimed to identify equivalent non-local continuum models of integral-differential and gradient type, the latter obtained through standard or enhanced continualization. This result has been achieved by imposing the equivalence of the frequency band structure of the discrete system with that one of the equivalent continuum. The bilateral Z-transform in the space of nodal displacements has been applied to the equation of motion of the Lagrangian model. After a remapping on the unit circle, the transformed equation takes the form of a continuous field equation in the domain of both the time and of the wave vectors. A suitable macro-field (i.e. macro-displacement) has been introduced, whose spatial Fourier transform is imposed to match the bilateral Z-transform mapped on the unit circle of the nodal displacements. Hence, the integral-differential field equation of a non-local continuum involving the macro-displacement field has been obtained. It has been shown that the approximation of the kernel of such integral-differential equation through truncated power series leads to differential field equations of higher order representative of continua with local or non-local gradient in the constitutive part only. Moreover, it has been shown that this differential governing equation corresponds to the ones obtained through the standard continualization.

Energetically consistent equivalent continua have been identified through a proper mapping correlating the transformed macro-displacements in the Fourier space with a new auxiliary regularizing continuum macro-displacement field in the same space. Specifically, the mapping here introduced has zeros at the edge of the first Brillouin zone. Then, the integral-differential governing equation and the corresponding differential one have been reformulated, which are characterized by both inertial and constitutive non-locality. In particular, the constitutive and inertial kernels of the integral-differential equation exhibit polar singularities at the edge of the first Brillouin zone.

This procedure appears to agree with the enhanced continualization of one-dimensional Lagrangian systems proposed by Bacigalupo and Gambarotta (2019) whose formulation was based on shift and pseudo-differential operators. Here, the frequency spectrum of the gradient continuum obtained through the enhanced continualization has been found to converge to the actual one, namely to that of the integral-differential continuum, as the order of the gradient continuum increases. This result is justified since the convergence radius of the power series expansion of the constitutive and inertial kernels turns out to equal the half of the characteristic size of the first Brillouin zone.

The proposed approach has been consistently generalized to the two-dimensional case by using multidimensional bilateral Z- and Fourier transforms and an appropriate generalization of the auxiliary regularized continuous macro-displacement field. This procedure may be easily extended to three-dimensional lattices. Finally, two examples of lattice-like systems consisting of periodic pre-stressed cable-net of mass points are analyzed. In these examples it is shown as the gradient continuum models characterized by differential equations obtained as approximations of the integral-differential ones have dispersion curves who are in excellent agreement with those of the Lagrangian model, converging to that one of the discrete system when increasing the order of the gradient continuum.

**Acknowledgements**

The authors acknowledge financial support of the (MURST) Italian Department for University and Scientific and Technological Research in the framework of the research MIUR Prin15 project 2015LYYXA8, Multi-scale mechanical models for the design and optimization of micro-structured smart materials and metamaterials, coordinated by prof. A. Corigliano. AB thankfully acknowledge financial support by National Group of Mathematical Physics (GNFM-INdAM).

**Appendix A**

From definition (6), where the transformed displacement field at the macroscale is assumed to coincide with the Z-transform of the nodal displacement mapped on the unit-circle, i.e.

$$\hat{\Psi}(k,t)=\mathcal{F}\left[\Psi(x,t)\right]\doteq\mathcal{Z}\left[\psi_i\right]\Big|_{z=\exp(Ik\ell)}, \tag{82}$$

and by rearranging the mapping (17) in the transformed space

$$2Ik\ell\,\hat{\Psi}^R(k,t)=\left[\exp(Ik\ell)-\exp(-Ik\ell)\right]\hat{\Psi}(k,t) \tag{83}$$

one gets after an inverse Fourier transform

$$\mathcal{F}^{-1}\left[2Ik\ell\,\mathcal{F}\left[\Psi^R(x,t)\right]\right]=\mathcal{F}^{-1}\left[\left(\exp(Ik\ell)-\exp(-Ik\ell)\right)\,\mathcal{F}\left[\Psi(x,t)\right]\right]. \tag{84}$$

It is worth to note that equation (84) may be specialized as

$$\begin{aligned}\frac{\partial\Psi^R}{\partial x}&=\frac{1}{2\ell}\mathcal{F}^{-1}\left[\left(\exp(Ik\ell)-\exp(-Ik\ell)\right)\,\mathcal{F}\left[\Psi(x,t)\right]\right]=\\&=\frac{1}{2\pi}\int_{-\infty}^{+\infty}\frac{\exp(Ik\ell)-\exp(-Ik\ell)}{2\ell}\mathcal{F}\left[\Psi(x,t)\right]\exp(Ikx)\,\ell dk\end{aligned} \tag{85}$$

that, by introducing the concept of pseudo-differential operator, takes the form

$$D\Psi^R(x,t)=\left[\frac{\exp(\ell D)-\exp(-\ell D)}{2\ell}\right]\Psi(x,t) \tag{86}$$

in agreement with Bacigalupo and Gambarotta (2019).

From equation (85) and recalling (82) one has

$$\begin{aligned}\frac{\partial\Psi^R}{\partial x}&=\frac{1}{2\pi}\int_{-\infty}^{+\infty}\frac{\exp(Ik\ell)-\exp(-Ik\ell)}{2\ell}\mathcal{Z}\left[\psi_i(x,t)\right]\Big|_{z=\exp(Ik\ell)}\exp(Ikx)\,\ell dk=\\&=\frac{1}{2\pi}\int_{-\infty}^{+\infty}\frac{1}{2\ell}\left(Z\left[\psi_{i+1}(x,t)\right]\Big|_{z=\exp(Ik\ell)}-\mathcal{Z}\left[\psi_{i-1}(x,t)\right]\Big|_{z=\exp(Ik\ell)}\right)\exp(Ikx)\,\ell dk\end{aligned} \tag{87}$$

and by remembering the Z-transform definition equation (87) may be rearranged as

$$\frac{\partial \Psi^R}{\partial x} = \frac{1}{2\ell}\begin{bmatrix} \frac{1}{2\pi}\int_{-\infty}^{+\infty} \exp(Ikx)\sum_{-\infty}^{+\infty}\psi_j \exp(-Ijk\ell)\exp(Ik\ell)\,\ell dk + \\ -\frac{1}{2\pi}\int_{-\infty}^{+\infty} \exp(Ikx)\sum_{-\infty}^{+\infty}\psi_j \exp(-Ijk\ell)\exp(-Ik\ell)\,\ell dk \end{bmatrix} = \\ = \frac{1}{2\ell}\begin{bmatrix} \frac{1}{2\pi}\sum_{-\infty}^{+\infty}\psi_j\int_{-\infty}^{+\infty} \exp\left[Ik\left(x+\ell(1-j)\right)\right]\ell dk + \\ -\frac{1}{2\pi}\sum_{-\infty}^{+\infty}\psi_j\int_{-\infty}^{+\infty} \exp\left[Ik\left(x-\ell(1+j)\right)\right]\ell dk + \end{bmatrix}. \tag{88}$$

This derivative evaluated at the characteristic points of the lattice $x = n\ell = x_n$ with $n \in \mathbb{Z}$ one has

$$\left.\frac{\partial \Psi^R}{\partial x}\right|_{x_n} = \frac{1}{2\ell}\begin{bmatrix} \frac{1}{2\pi}\sum_{-\infty}^{+\infty}\psi_j\int_{-\infty}^{+\infty} \exp\left[Ik\left(n+\ell(1-j)\right)\right]\ell dk + \\ -\frac{1}{2\pi}\sum_{-\infty}^{+\infty}\psi_j\int_{-\infty}^{+\infty} \exp\left[Ik\left(n-\ell(1+j)\right)\right]\ell dk + \end{bmatrix} = \\ = \frac{1}{2\ell}\left[\frac{1}{2\pi}\sum_{-\infty}^{+\infty}2\pi\psi_j\,\delta_{jn+1} - \frac{1}{2\pi}\sum_{-\infty}^{+\infty}2\pi\psi_j\,\delta_{jn-1}\right] = \\ = \frac{\psi_{n+1}-\psi_{n-1}}{2\ell}\,, \tag{89}$$

being $\delta_{jn+1}$ and $\delta_{jn-1}$ the Kronecker delta functions. An equivalent form of equation (89) may be derived

$$\left.\frac{\partial \Psi^R}{\partial x}\right|_{x_n} = \frac{\psi_{n+1}-\psi_{n-1}}{2\ell} = \left[\frac{\exp(\ell D)-\exp(-\ell D)}{2\ell}\right]\psi_n, \tag{90}$$

where $\exp(\ell D)$ plays the role of shift operator according to Maslov (1976), Shubin (1987) and Bacigalupo and Gambarotta (2019).

## Appendix B

The real and imaginary part of the constitutive kernel $\mathcal{K}_2(k\ell)$ involved in equation (20) take the form

$$\Re e(\mathcal{K}_2) = \frac{\begin{bmatrix} 2\,\Im m(k\ell)\cosh(\Im m(k\ell))\sinh(\Im m(k\ell)) - 2\,\Re e(k\ell)\cosh(\Im m(k\ell))\sin(\Re e(k\ell)) + \\ -2\Im m(k\ell)\cos(\Re e(k\ell))\sinh(\Im m(k\ell)) + 2\Re e(k\ell)\cos(\Re e(k\ell))\sin(\Re e(k\ell)) \end{bmatrix}}{\left(\cosh(\Im m(k\ell))\right)^2 - \left(\cos(\Re e(k\ell))\right)^2}, \tag{91}$$

$$\Im m\left(\mathcal{K}_2\right)=\frac{\begin{bmatrix}2\Re e(k\ell)\cos\left(\Re e(k\ell)\right)\sinh\left(\Im m(k\ell)\right)-2\Im m(k\ell)\cosh\left(\Im m(k\ell)\right)\sin\left(\Re e(k\ell)\right)+\\ -2\,\Re e(k\ell)\cosh\left(\Im m(k\ell)\right)\sinh\left(\Im m(k\ell)\right)+2\,\Im m(k\ell)\cos\left(\Re e(k\ell)\right)\sin\left(\Re e(k\ell)\right)\end{bmatrix}}{\left(\cosh\left(\Im m(k\ell)\right)\right)^2-\left(\cos\left(\Re e(k\ell)\right)\right)^2},\tag{92}$$

and the magnitude reads

$$\left|\mathcal{K}_2\right|=2\left|k\ell\right|\sqrt{\frac{\left(\cosh\left(\Im m(k\ell)\right)-\cos\left(\Re e(k\ell)\right)\right)}{\cosh\left(\Im m(k\ell)\right)+\cos\left(\Re e(k\ell)\right)}}\,.\tag{93}$$

Finally, the real and imaginary part of the inertial kernel $\mathcal{K}_3\left(k\ell\right)$ involved in equation (20) take the form

$$\Re e\left(\mathcal{K}_3\right)=\frac{\Re e(k\ell)\sin\left(\Re e(k\ell)\right)\cosh\left(\Im m(k\ell)\right)+\Im m(k\ell)\cos\left(\Re e(k\ell)\right)\sinh\left(\Im m(k\ell)\right)}{\left(\cosh\left(\Im m(k\ell)\right)\right)^2-\left(\cos\left(\Re e(k\ell)\right)\right)^2},\tag{94}$$

$$\Im m\left(\mathcal{K}_3\right)=\frac{\Im m(k\ell)\sin\left(\Re e(k\ell)\right)\cosh\left(\Im m(k\ell)\right)-\Re e(k\ell)\cos\left(\Re e(k\ell)\right)\sinh\left(\Im m(k\ell)\right)}{\left(\cosh\left(\Im m(k\ell)\right)\right)^2-\left(\cos\left(\Re e(k\ell)\right)\right)^2},\tag{95}$$

and the magnitude reads

$$\left|\mathcal{K}_3\right|=\frac{2\left|k\ell\right|}{\sqrt{\left(\cosh\left(\Im m(k\ell)\right)\right)^2+\left(\cos\left(\Re e(k\ell)\right)\right)^2}}\,.\tag{96}$$